\documentclass[usenatbib]{mnras}

\pdfoutput=1

\usepackage[utf8]{inputenc}
\usepackage[pdftex]{graphicx}
\usepackage{mathptmx}
\usepackage{fixltx2e}
\usepackage{color}
\usepackage{amssymb}

\usepackage{hyperref}

\usepackage{etoolbox}
\makeatletter
\newcount\c@additionalboxlevel
\newcount\c@maxboxlevel
\patchcmd\@combinedblfloats{\box\@outputbox}{%
  \stepcounter{additionalboxlevel}%
  \box\@outputbox
}{}{\errmessage{\noexpand\@combinedblfloats could not be patched}}

\AtBeginShipout{%
  \ifnum\value{additionalboxlevel}>\value{maxboxlevel}%
    \typeout{Warning: maxboxlevel might be too small, increase to %
      \the\value{additionalboxlevel}%
    }%
  \fi
  \@whilenum\value{additionalboxlevel}<\value{maxboxlevel}\do{%
    \typeout{* Additional boxing of page `\thepage'}%
    \setbox\AtBeginShipoutBox=\hbox{\copy\AtBeginShipoutBox}%
    \stepcounter{additionalboxlevel}%
  }%
  \setcounter{additionalboxlevel}{0}%
}
\makeatother

\newcommand{\kpc}{\rm\thinspace kpc}

\newcommand{\cm}{\rm\thinspace cm}
\newcommand{\s}{\rm\thinspace s}

\newcommand{\photon}{\rm photon}
\newcommand{\pixel}{\rm pixel}

\newcommand{\photoncmsqps}{\hbox{$\photon\cm^{-2}\s^{-1}\,$}}

\newcommand{\Zsun}{\hbox{$\thinspace \mathrm{Z}_{\odot}$}}

\newcommand{\pcmsq}{\hbox{$\cm^{-2}\,$}}

\begin{document}

\title[Galaxy cluster edge detection]
{Detecting edges in the X-ray surface brightness of galaxy clusters}

\author
[J.~S. Sanders et al.]
{
  \begin{minipage}[b]{\linewidth}
    \begin{flushleft}
      J.~S.~Sanders,$^1$\thanks{E-mail: jsanders@mpe.mpg.de}
      A.~C.~Fabian,$^2$
      H.~R.~Russell,$^2$
      S.~A.~Walker,$^2$ and
      K.~M.~Blundell$^3$
    \end{flushleft}
  \end{minipage}
  \\
  $^1$ Max-Planck-Institut f\"ur extraterrestrische Physik,
  Giessenbachstrasse 1, D-85748 Garching, Germany\\
  $^2$ Institute of Astronomy, Madingley Road, Cambridge, CB3 0FT\\
  $^3$ University of Oxford, Astrophysics, Keble Road, Oxford, OX1 3RH\\
}
\maketitle

\begin{abstract}
  The effects of many physical processes in the intracluster medium of
  galaxy clusters imprint themselves in X-ray surface brightness
  images. It is therefore important to choose optimal methods for
  extracting information from and enhancing the interpretability of
  such images. We describe in detail a gradient filtering edge
  detection method that we previously applied to images of the
  Centaurus cluster of galaxies. The Gaussian gradient filter measures
  the gradient in the surface brightness distribution on particular
  spatial scales. We apply this filter on different scales to
  \emph{Chandra X-ray observatory} images of two clusters with AGN
  feedback, the Perseus cluster and M\,87, and a merging system,
  A\,3667. By combining filtered images on different scales using
  radial filters spectacular images of the edges in a cluster are
  produced. We describe how to assess the significance of features in
  filtered images. We find the gradient filtering technique to have
  significant advantages for detecting many kinds of features compared
  to other analysis techniques, such as unsharp-masking. Filtering
  cluster images in this way in a hard energy band allows shocks to be
  detected.
\end{abstract}


\section{Introduction}
X-ray emission from clusters is mainly due to bremsstrahlung emission
\citep{Felten66,Mitchell76} from the hot intracluster medium
(ICM). The X-ray flux is proportional to the square of density, with
some temperature dependence, and so is a sensitive tracer of
variations in the thermodynamic properties of the ICM.

Although relaxed clusters are largely hot atmospheres in hydrostatic
pressure equilibrium, density and temperature variations are important
probes of astrophysical processes within the cluster and of cluster-wide
perturbations such as mergers. AGN feedback in clusters
\citep{Fabian12} injects bubbles of radio plasma into the ICM,
displacing the X-ray emitting gas and creating cavities in X-ray
images \citep[e.g.][]{BohringerPer93,FabianPer00,McNamara00}. In
addition, AGN are observed to shock their surroundings
\citep[e.g.][]{Forman07,Randall15} and likely generate sound waves in
the ICM \citep{FabianPer06,SandersSound08,Blanton11}. These processes
are seen in simulations of AGN feedback
\citep[e.g.][]{Ruszkowski04,Sijacki06}.

Cold fronts --- discontinuities in temperature and density --- are
extremely common in clusters \citep{MarkevitchCFShock07}. These are
seen in both merging \citep[e.g.][]{Vikhlinin01} and relaxed
\citep[e.g.][]{PaternoMahler13} clusters. In the relaxed cases the fronts are
believed to be caused by gas sloshing in the potential well due to the
passage of a subcluster \citep{Ascasibar06}. Such sloshing may remain
for several Gyr and can give rise to several edges within a single
cluster. Merging subclusters additionally generate shocks, also seen
as surface brightness edges \citep[e.g.][]{Markevitch02,Russell10}.

Mergers and AGN feedback can inject turbulence in the ICM
\citep{NormanBryan99}. Turbulence within the ICM should be associated
with gas density and therefore surface brightness variations which can
be quantified
\citep{Churazov12,SandersAWM712,ZhuravlevaFluct14,Walker15}.

The radial variation in density in clusters, particularly in relaxed
objects, gives rise to steeply peaked surface brightness
profiles. Therefore it is often difficult to see variations in surface
brightness profile on these mountainous X-ray peaks.  Various
techniques have been developed to enhance the structure seen in images
of galaxy clusters, allowing variations to be observed. These include
dividing or subtracting a symmetric model, such as the $\beta$ model
\citep[e.g.][]{Arnaud01}, or the average at each radius
\citep[e.g.][]{Churazov99}. More complex models can also be used, such
as fits to surface brightness contours with ellipses
\citep{SandersAWM712}.

Another common technique is to use unsharp masking of cluster images
to highlight smaller-scale substructure
\citep[e.g.][]{FabianPer03}. The typical method is to smooth the image
first by a Gaussian with a large width and subtract (or divide) this
from the same image smoothed by a Gaussian with a smaller width. The
process results in an image which suppresses both large and small
scale structure. A further method to suppress
unwanted cluster signal
is a Fourier bandpass filter \citep{SandersSound08}. However, a risk
with Fourier techniques is that spurious circular ringing artifacts
from edges and point sources can be introduced if frequency cut-offs
in the applied filter are too sharp.

Here we apply a gradient measuring filter to X-ray images of galaxy
clusters. Finding edges and measuring gradients in surface brightness
are useful, because all the astrophysical processes we have previously
mentioned introduce density variations, and therefore surface
brightness gradients, into X-ray images. For example, shocks and cold
fronts produce edges (i.e. very steep gradients), while sound waves
should produce alternating flat and steep gradients. As we are not
interested in the total X-ray emission in a region, using the gradient
removes much of the X-ray peak. Gradient filtering has previously been
used in examining simulations of galaxy cluster to look for edges
associated with cold fronts and sloshing \citep{Roediger13}.

The use of the Gaussian Gradient Magnitude (GGM) filter was introduced
to X-ray analysis in our study of deep \emph{Chandra} observations of
the Centaurus cluster \citep{SandersCent16}.  The GGM filter
calculates the gradient of an image assuming Gaussian derivatives
(with a width $\sigma$). In comparison, the Sobel operator,
a type of
gradient filter, convolves two $3\times3$ matrices with the image and
can be used to compute the magnitude of the gradient in an image on a
pixel-by-pixel basis.  The advantage of the GGM filter over the Sobel
filter is that $\sigma$ can be adjusted to measure the gradient over
more or fewer pixels depending on the data quality and the scale of
the features of interest. In X-ray cluster images, a large $\sigma$
would be used for regions in the outskirts where there are few or no
X-ray counts per pixel and a small value in the centre where there are
many counts.

In this work we apply the GGM filter to other high quality data sets
from the \emph{Chandra} archive to demonstrate that the technique is a
powerful method for the identification of physical processes taking
place in the ICM. We examine two relaxed clusters with short central
cooling times and active AGN feedback and one disturbed system
undergoing a merger. In the Perseus cluster, A\,426, there are
multiple X-ray cavities \citep{Bohringer04,FabianPer00}, a weak shock
\citep{FabianPer03}, ripples \citep{FabianPer06} and uplifted high
metallicity material \citep{SandersNonTherm05}, which may be sound
waves from AGN feedback. M\,87 contains a bright jet, multiple
bubble-like cavities \citep{Young02} and a weak shock
\citep{Forman05,Forman07}. There are cool arms of metal rich material
being dragged out by the radio bubbles
\citep{Young02,Simionescu07,Million10}. A\,3667 is a system undergoing a
merger \citep{Knopp96} and hosts a sharp surface brightness
discontinuity, a cold front, indicating material is moving through the
ambient gas with a Mach number of $\sim 1$
\citep{Vikhlinin01}. Optically, the cluster has two distinct sets of
galaxies \citep{Sodre92} and a radio relic \citep{Rottgering97}.

\section{Data preparation}
\begin{table}
  \caption{\emph{Chandra} data sets examined for the Perseus cluster.
    For each observation we list the observation identifier, observation
    starting date, exposure (unfiltered and after filtering for flares in ks)
    and the ACIS CCDs from which the data were examined, with the ACIS mode
    (S or I). $^{*}$ marks the reference observations that others
    were reprojected to.}
  \centering
  \begin{tabular}{rrrrr}
    OBSID & Date & Exposure & Filtered & CCDs \\ \hline
      502 & 1999-09-20 &   5.1 &   2.4 &  I: 0,1,2,3,6,7 \\
      503 & 1999-11-28 &   9.0 &   8.8 &  S: 2,3,6,7,8 \\
     1513 & 2000-01-29 &  24.9 &  10.5 &  S: 2,3,6,7,8 \\
     3209 & 2002-08-08 &  95.8 &  94.0 &  S: 1,3,6,7 \\
     4289 & 2002-08-10 &  95.4 &  93.0 &  S: 1,3,6,7 \\
     6139 & 2004-10-04 &  56.4 &  53.2 &  S: 2,3,5,6,7 \\
     4946 & 2004-10-06 &  23.7 &  23.3 &  S: 2,3,5,6,7 \\
     4948 & 2004-10-09 & 118.6 & 111.3 &  S: 2,3,5,6,7 \\
     4947 & 2004-10-11 &  29.8 &  29.4 &  S: 2,3,5,6,7 \\
     4949 & 2004-10-12 &  29.4 &  29.2 &  S: 2,3,5,6,7 \\
     4950 & 2004-10-12 &  96.9 &  75.6 &  S: 2,3,5,6,7 \\
$^{*}$4952& 2004-10-14 & 164.2 & 147.3 &  S: 2,3,5,6,7 \\
     4951 & 2004-10-17 &  96.1 &  93.8 &  S: 2,3,5,6,7 \\
     4953 & 2004-10-18 &  30.1 &  29.7 &  S: 2,3,5,6,7 \\
     6145 & 2004-10-19 &  85.0 &  84.6 &  S: 2,3,5,6,7 \\
     6146 & 2004-10-20 &  47.1 &  42.4 &  S: 2,3,5,6,7 \\
    11716 & 2009-10-10 &  39.6 &  38.1 &  I: 0,1,2,3,6 \\
    12025 & 2009-11-25 &  17.9 &  11.7 &  I: 0,1,2,3,6,7 \\
    12033 & 2009-11-27 &  18.9 &  12.0 &  I: 0,1,2,3,6,7 \\
    11713 & 2009-11-29 & 112.2 &  75.9 &  I: 0,1,2,3,6,7 \\
    12036 & 2009-12-02 &  47.9 &  34.8 &  I: 0,1,2,3,6,7 \\
    11715 & 2009-12-02 &  73.4 &  68.6 &  I: 0,1,2,3,6 \\
    12037 & 2009-12-05 &  84.6 &  79.1 &  I: 0,1,2,3,6 \\
    11714 & 2009-12-07 &  92.0 &  80.1 &  I: 0,1,2,3,6 \\ \hline
    Total &            &       &1328.7 &              \\
  \end{tabular}
  \label{tab:per_obs}
\end{table}

\begin{table}
  \caption{\emph{Chandra} data sets examined for M\,87.
    The columns are the same as for Table \ref{tab:per_obs}.}
  \centering
  \begin{tabular}{rrrrr}
    OBSID & Date & Exposure & Filtered & CCDs \\ \hline
      352 & 2000-07-29 &  37.7 &  33.5 & S: 2,3,6,7 \\
     3717 & 2002-07-05 &  20.6 &  10.5 & S: 2,3,6,7 \\
$^{*}$2707& 2002-07-06 &  98.7 &  88.2 & S: 2,3,6,7 \\
     6186 & 2005-01-31 &  51.6 &  42.8 & I: 0,1,2,3 \\
     5826 & 2005-03-03 & 126.8 & 120.0 & I: 0,1,2,3 \\
     5827 & 2005-05-05 & 156.2 & 147.7 & I: 0,1,2,3 \\
     7212 & 2005-11-14 &  65.2 &  61.2 & I: 0,1,2,3 \\
     7210 & 2005-11-16 &  30.7 &  27.5 & I: 0,1,2,3 \\
     7211 & 2005-11-16 &  16.6 &  15.7 & I: 0,1,2,3 \\
     5828 & 2005-11-17 &  33.0 &  31.4 & I: 0,1,2,3 \\
    15180 & 2013-08-01 & 138.8 & 137.2 & I: 0,1,2,3 \\
    15178 & 2014-02-17 &  46.5 &  46.1 & I: 0,1,2,3 \\
    16585 & 2014-02-19 &  45.0 &  44.0 & I: 0,1,2,3 \\
    16586 & 2014-02-20 &  49.2 &  48.4 & I: 0,1,2,3 \\
    16587 & 2014-02-22 &  37.3 &  37.0 & I: 0,1,2,3 \\
    15179 & 2014-02-24 &  41.4 &  40.2 & I: 0,1,2,3 \\
    16590 & 2014-02-27 &  37.6 &  37.0 & I: 0,1,2,3 \\
    16591 & 2014-02-27 &  23.5 &  22.9 & I: 0,1,2,3 \\
    16592 & 2014-03-01 &  35.6 &  35.0 & I: 0,1,2,3 \\
    16593 & 2014-03-02 &  37.6 &  36.8 & I: 0,1,2,3 \\ \hline
    Total &            &       &1063.2 &            \\
  \end{tabular}
  \label{tab:m87_obs}
\end{table}

\begin{table}
  \caption{\emph{Chandra} data sets examined for A\,3667.
    The columns are the same as for Table \ref{tab:per_obs}.}
  \centering
  \begin{tabular}{rrrrr}
    OBSID & Date & Exposure & Filtered &  CCDs          \\ \hline
      513 & 1999-09-22 &  44.8 &  40.1 & I: 0,1,2,3,5,6 \\
      889 & 2000-09-09 &  50.3 &  49.7 & I: 0,1,2,3,6,7 \\
$^{*}$5751& 2005-06-07 & 128.9 & 125.0 & I: 0,1,2,3,6   \\
     6292 & 2005-06-10 &  46.7 &  45.7 & I: 0,1,2,3,6   \\
     5752 & 2005-06-12 &  60.4 &  59.4 & I: 0,1,2,3,6   \\
     6295 & 2005-06-15 &  49.5 &  48.9 & I: 0,1,2,3,6   \\
     5753 & 2005-06-17 & 103.6 &  74.0 & I: 0,1,2,3,6   \\
     6296 & 2005-06-19 &  49.4 &  48.6 & I: 0,1,2,3,6   \\
     7686 & 2007-06-23 &   5.0 &   5.0 & I: 0,1,2,3     \\ \hline
    Total &            &       & 496.5 &                \\
  \end{tabular}
  \label{tab:a3667_obs}
\end{table}

\begin{figure*}
  \includegraphics[width=\textwidth]{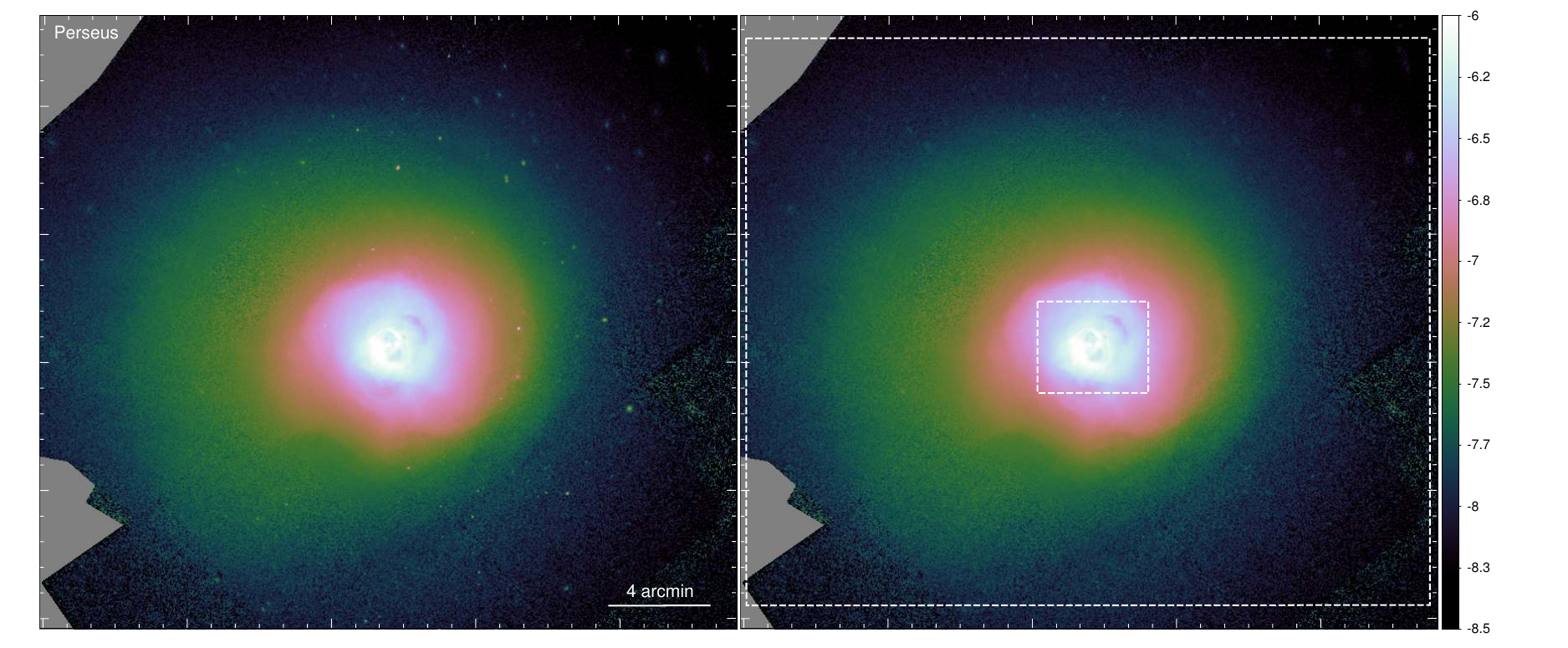}
  \includegraphics[width=\textwidth]{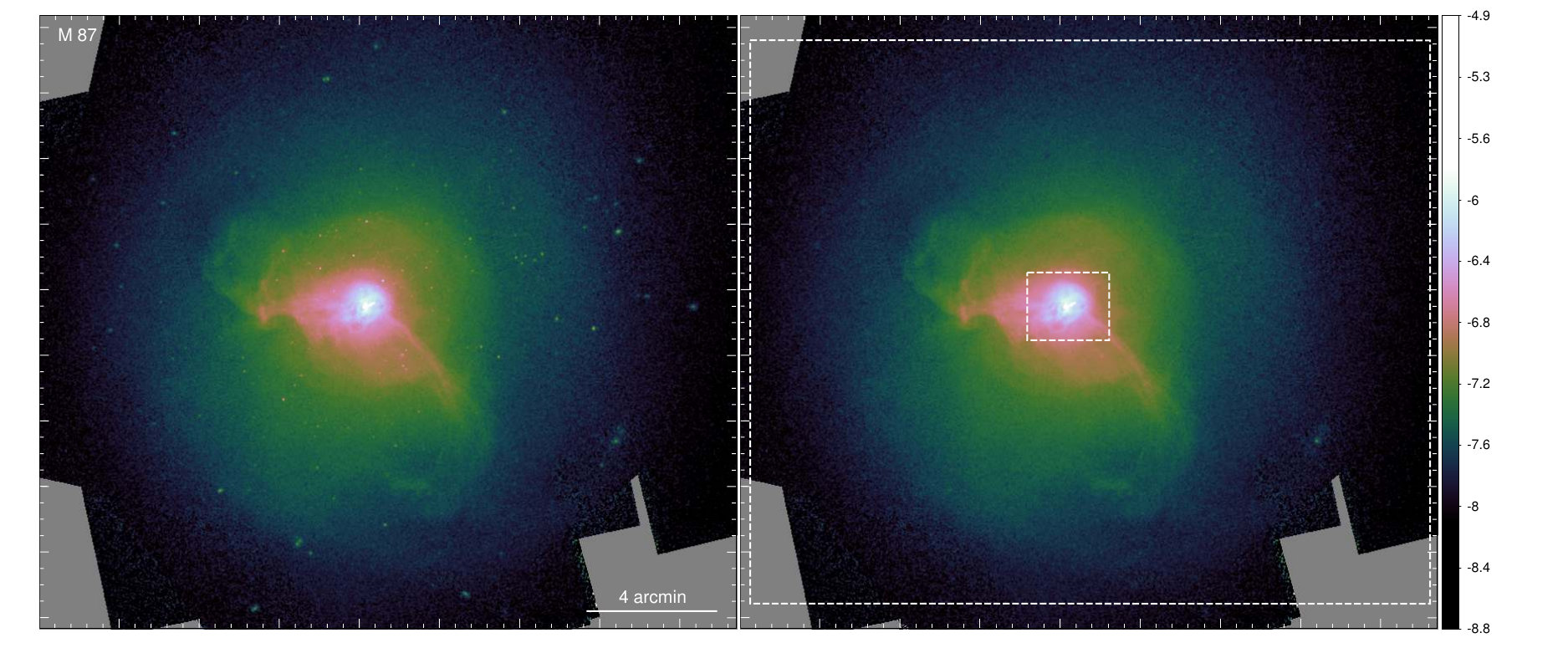}
  \includegraphics[width=\textwidth]{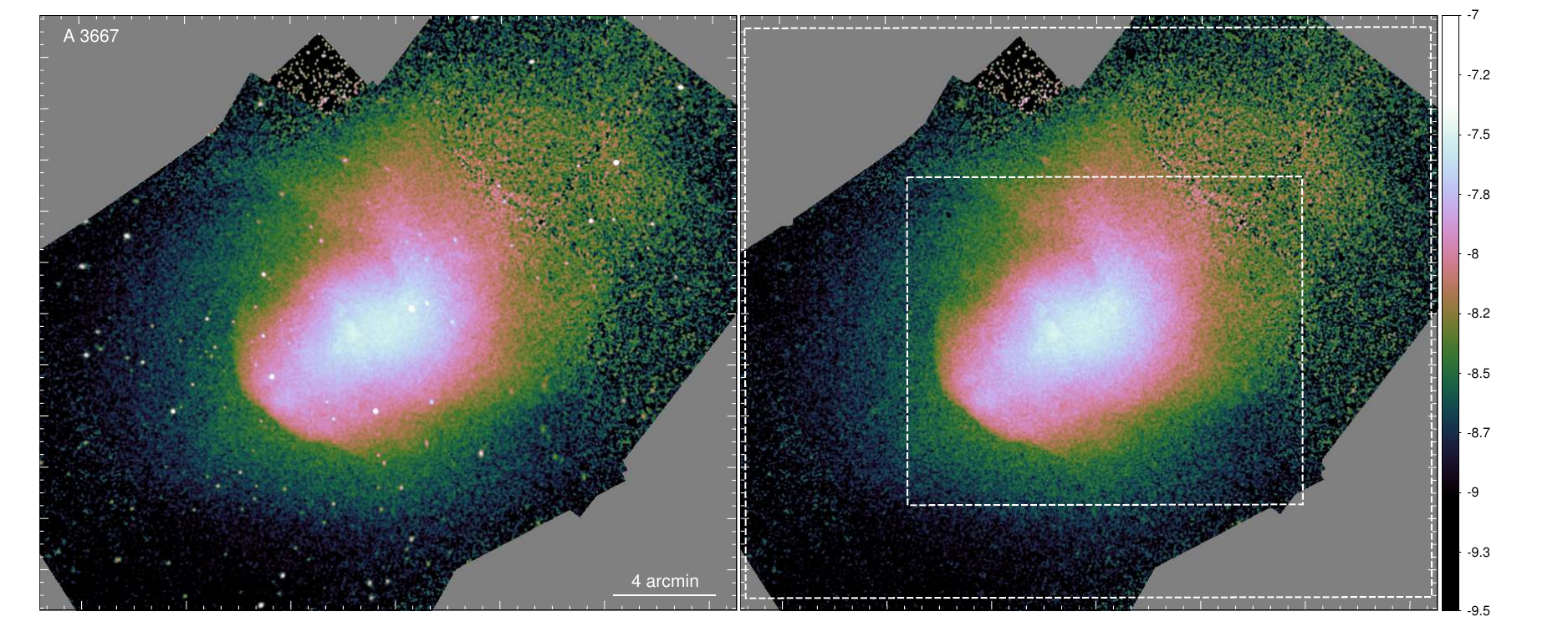}
  \caption{Exposure-corrected background-subtracted images of the
    Perseus cluster (top row), M\,87 (centre row) and A\,3667 (bottom
    row) in the 0.5 to 7 keV band. The left panels include the point
    sources, while they are cosmetically hidden in the right
    panels. The colour bar units are $\log_{10} \photoncmsqps
    \pixel^{-1}$. The Perseus and M\,87 images were smoothed by a
    Gaussian of 2 pixels (0.984 arcsec), while A\,3667 was smoothed by
    4 pixels. North is to the top and east is to the right in all the
    images in this paper. The white dashed boxes
      show the range in areas shown in the individual scale images.}
  \label{fig:expcorr}
\end{figure*}

We downloaded the data sets listed in Tables \ref{tab:per_obs},
\ref{tab:m87_obs} and \ref{tab:a3667_obs} from the \emph{Chandra}
archive. The data sets were reprocessed using
\textsc{ciao} \citep{Fruscione06} version 4.7
and \textsc{caldb} version 4.4.10. We filtered bad time periods using
X-ray lightcurves. For observations taken with ACIS-S (the
S subarray on the Advanced CCD Imaging Spectrometer), we extracted lightcurves from
CCD 5, if available, otherwise CCD 7. When analysing ACIS-I data, we
used CCDs 0, 1 and 2. The bad time periods were chosen using an
iterative $\sigma$ clipping algorithm, clipping 200s periods with rates
outside $2.5\sigma$ where $\sigma$ is the Poisson error on the mean
count rate.  The individual observations were reprojected to the
coordinate system of the marked reference observations,
indicated
with a $^{*}$ in Tables \ref{tab:per_obs}, \ref{tab:m87_obs} and
\ref{tab:a3667_obs}. Images were constructed for each
observation and CCD using single-detector pixel binning.

Spatial masks were applied to the data to improve the quality
of the output images. We excluded the outer edges of
the ACIS-I array and those of individual ACIS-S CCDs. It is unclear why
these steps should be necessary, but doing these removed structures
associated with the edges. For M\,87, we also applied narrow exclusion
regions along the CCD read-out direction to
remove the readout streak associated with out-of-time events. In the
Perseus cluster, we masked out some regions which were far off-axis in
some observations but covered by other observations with much better
PSFs. For example, we masked out the inner few arcmin in the offset
ACIS-I observations which were covered by on-axis ACIS-S observations.

We created exposure maps for each observation and CCD using
\textsc{mkexpmap}. As input spectra for the exposure map calculation,
we assumed for Perseus a $6$~keV plasma, with a metallicity of
$0.5\Zsun$, absorption equivalent to a Hydrogen column of $10^{21}
\pcmsq$ and a redshift of $0.0183$ \citep{SandersPer07}. For M\,87 we
assumed a temperature of $2.2$~keV, a metallicity of $1.1\Zsun$,
absorption of $1.93 \times 10^{20} \pcmsq$ \citep{Million10} and a
redshift of $0.004283$ \citep{Cappellari11}. For A\,3667 we took the
temperature to be $7$~keV and the ICM abundance to be $0.3\Zsun$
\citep{Vikhlinin01}, we used redshift of $0.0556$
\citep{StrubleRood99} and the Galactic column of $4.44 \times 10^{20}
\pcmsq$ \citep{Kalberla05}.

Standard blank-sky background event files were used to remove the
instrumental and diffuse X-ray background. For each foreground
observation we identified the appropriate backgrounds for each of the
CCDs. We filtered out events which fell on bad pixels of the
respective foreground observations. The exposure time of each
background dataset was adjusted to match the foreground rate in the 9
to 12 keV band. If the exposure times for the different CCD
backgrounds were different, we reduced the exposure times to match the
shortest background, discarding random events to keep the rate the
same. The backgrounds for each CCD were then merged and reprojected to
match the aspect of the respective foreground observations. We then
adjusted these observation backgrounds to have the same ratio of
exposure time to total background as the respective foreground
observation had to the total foreground exposure, by reducing exposure
times and discarding events appropriately. The observation backgrounds
were reprojected to the reference foreground observation. Background
images for each observation and CCD were made using the same binning
as the foreground images.

Total exposure-corrected and background-subtracted images
(Fig.~\ref{fig:expcorr} left panels) were then calculated from the
foreground, background and exposure-map images. Firstly, the spatial
masks were applied to each image. The total background image was
then subtracted from the total foreground image after scaling by the
ratio of exposure times. The total background-subtracted image was
then divided by the total exposure map.  We identified point sources
in the image by eye (the usual \textsc{wavdetect} detection tool
became confused by the high surface brightness central structures in
Perseus and M\,87). To cosmetically remove these from our exposure
corrected images, we replaced pixels inside the point source regions
with random values selected from the immediately surrounding
pixels. These cosmetically-corrected images can be seen in
Fig.~\ref{fig:expcorr} (right panels).

\section{Gradient filtering}
\subsection{Single scale filtering}
The GGM filter calculates the gradient of the image assuming Gaussian
derivatives. The gradient of the image is computed along the two axes
by convolving the image by the gradient of a one-dimensional Gaussian
function. These two gradient images are combined to create a total
gradient image. We used the implementation from \textsc{scipy}
(\url{http://scipy.org/}). As inputs we used the unsmoothed
exposure-corrected background-subtracted images, with the point
sources cosmetically removed.

\subsubsection{The Perseus cluster}
\begin{figure*}
  \includegraphics[width=\textwidth]{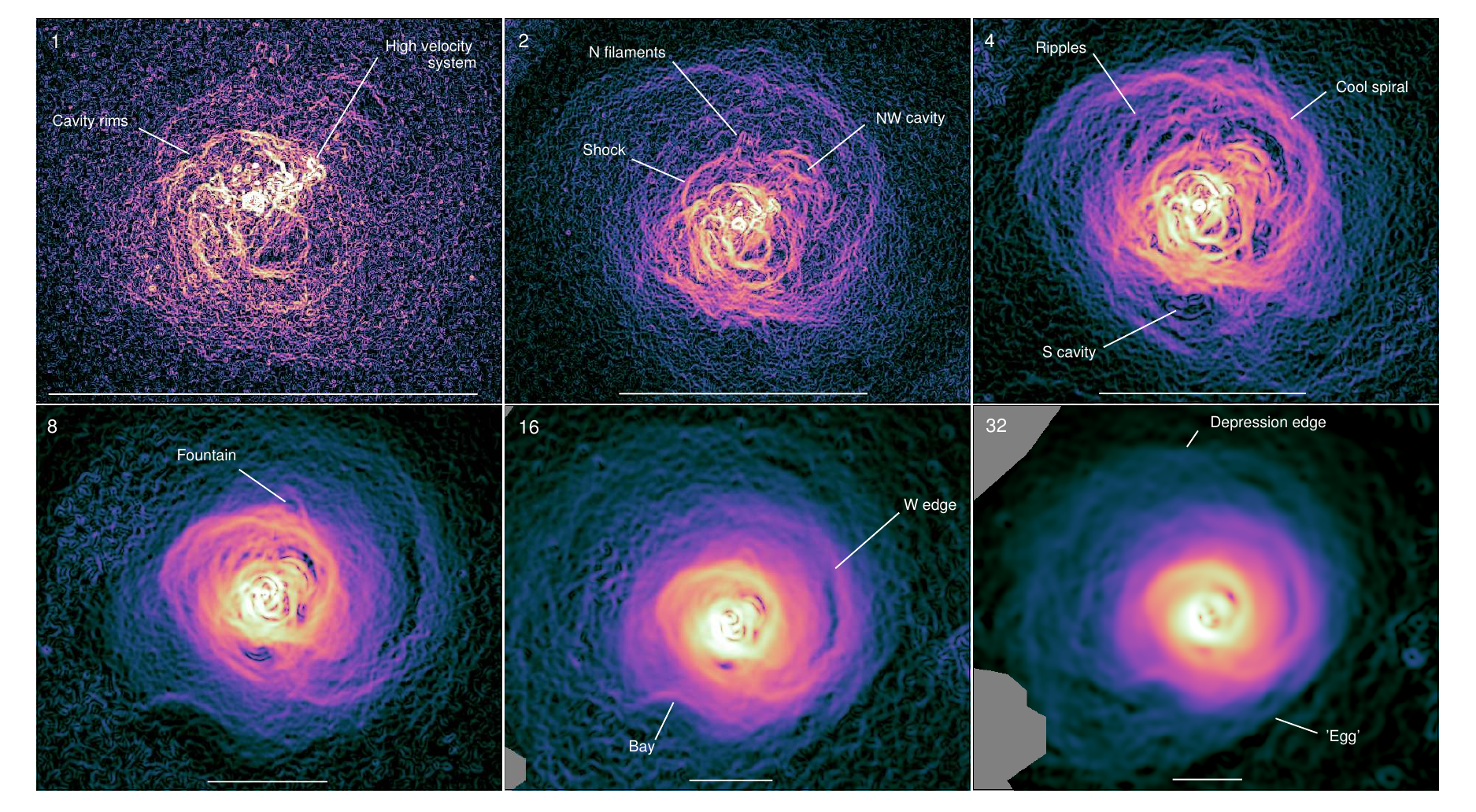}
  \caption{GGM-filtered images of the Perseus cluster with $\sigma=1$,
    2, 4, 8, 16 and 32 pixels (1 pixel is 0.492 arcsec). The bar has a
    length of 4 arcmin (89 kpc). The smallest and
      largest regions shown are indicated in Fig.~\ref{fig:expcorr}.}
  \label{fig:per_scales}
\end{figure*}

Fig.~\ref{fig:per_scales} shows filtered images of the Perseus cluster
using six exponentially increasing size scales. For each value of
$\sigma$ we show different regions of the cluster as it becomes harder
to measure gradients using small $\sigma$ when the count rate becomes
lower (see Section \ref{sect:detgrads}). Using
$\sigma=1$ and $2$, the image is sensitive to the finest structures in
the centre, which include the edges of the inner cavities and the
shocked region surrounding them \citep{FabianPer06}. Furthermore, some
of the features associated with the absorbing high-velocity system and
cool X-ray emitting filaments \citep{FabianPerFilament03} are
visible. Increasing the scales to $2$ and $4$, the cool spiral looping
around the north of the cluster \citep{SandersPer04} can be seen. To
the north is the `fountain', a structure seen in soft X-rays and by
its H$\alpha$ emission \citep{FabianPer06}. In addition the ripples in
X-ray surface brightness can be seen to the eastern side of the
cluster. In the 4 and 8 maps, the outer north-west and south ghost
cavities become visible, which are pointed to by low frequency radio
spurs \citep{FabianPer02}. Further out, in the 8 and 16 maps, is the
western edge \citep{Churazov03,FabianPer11}, curving in the same
direction as the inner spiral and at a radius of
  $\sim 110$~kpc. The top of the curve is flattened and is associated
with a surface brightness depression, one of a series in that
direction \citep{SandersPer07,FabianPer11}. These features may be
ancient relics of AGN activity, indicated by the radio emission and
long H$\alpha$ filament pointing towards them. In the 8 and 16 maps to
the south is the inverted-edge known as the `bay', which may be
another accumulation of previous AGN activity. On the largest scales,
the core of the cluster appears to be contained within an 9 arcmin
radius (200 kpc) egg-shaped region.

\begin{figure*}
  \includegraphics[width=\textwidth]{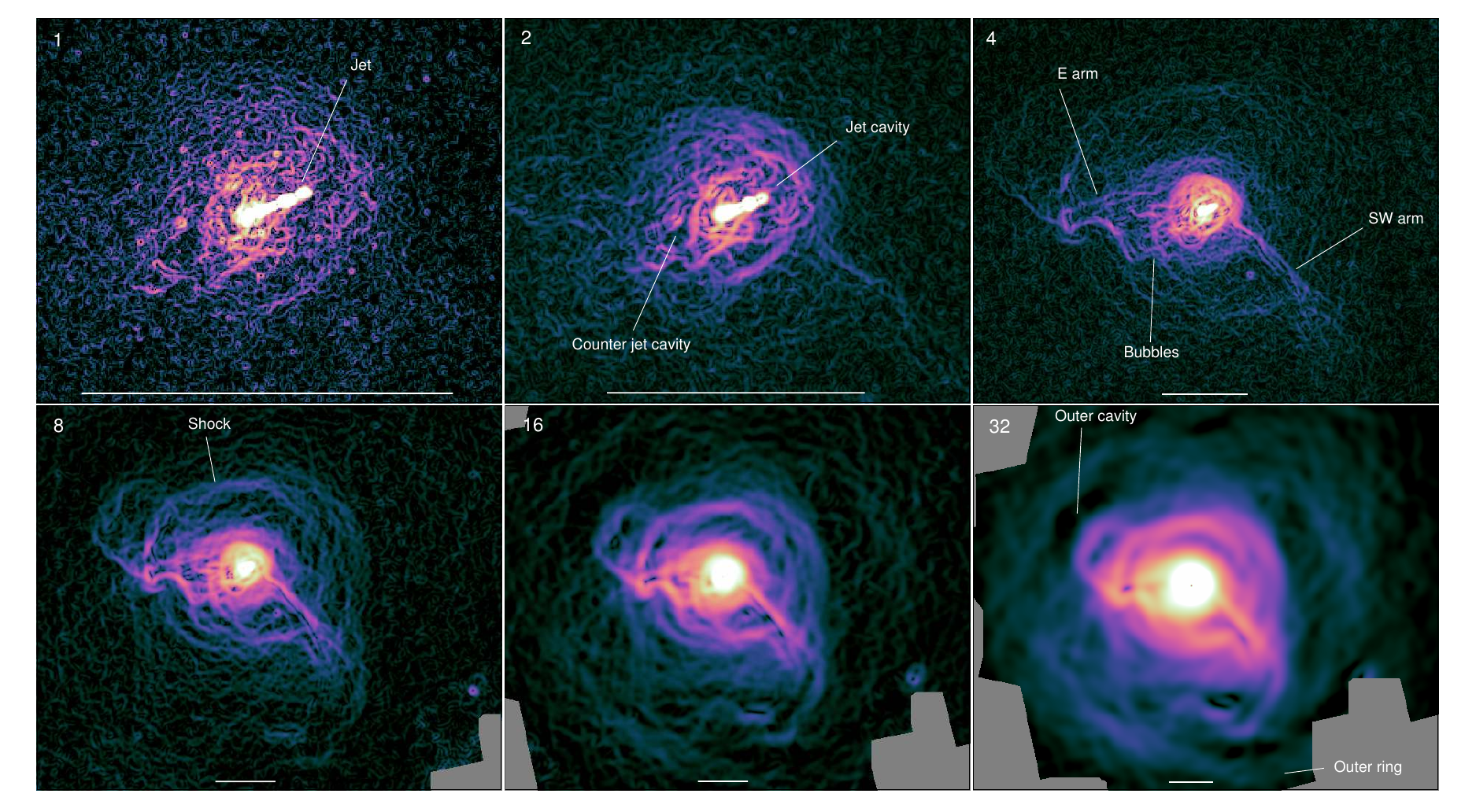}
  \caption{GGM-filtered images of M\,87 with $\sigma=1$, 2, 4, 8, 16
    and 32 pixels (1 pixel is 0.492 arcsec). The bar has a length of 2
    arcmin. Assuming a distance of 16.1 Mpc \citep{Tonry01}, this
    corresponds to 9.4 kpc.  The smallest and
      largest regions probed are shown in Fig.~\ref{fig:expcorr}.}
  \label{fig:m87_scales}
\end{figure*}

\begin{figure*}
  \includegraphics[width=\textwidth]{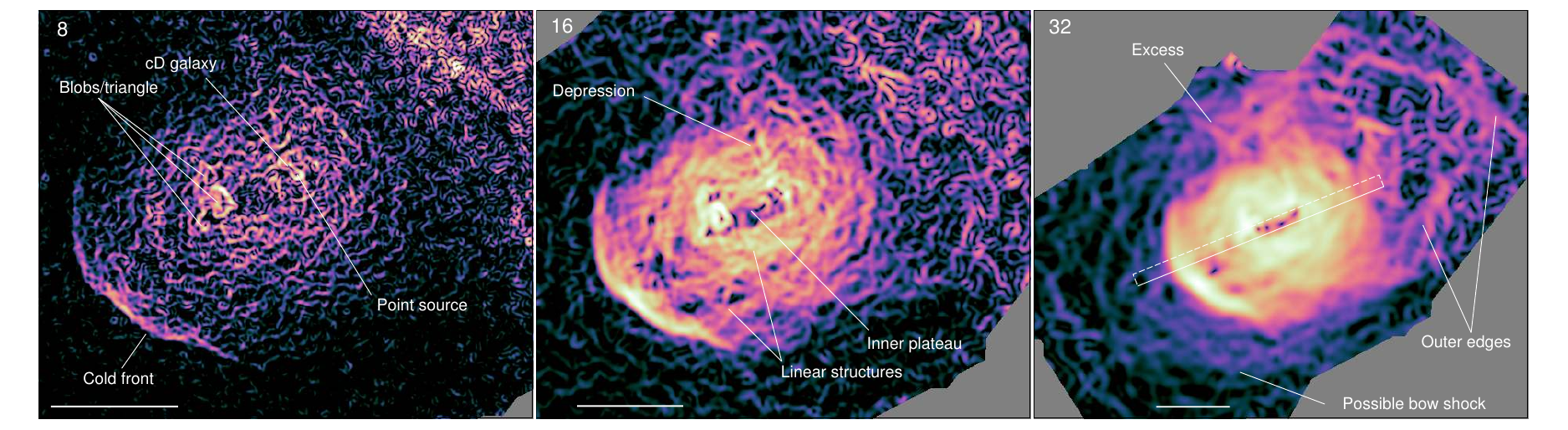}
  \includegraphics[width=\textwidth]{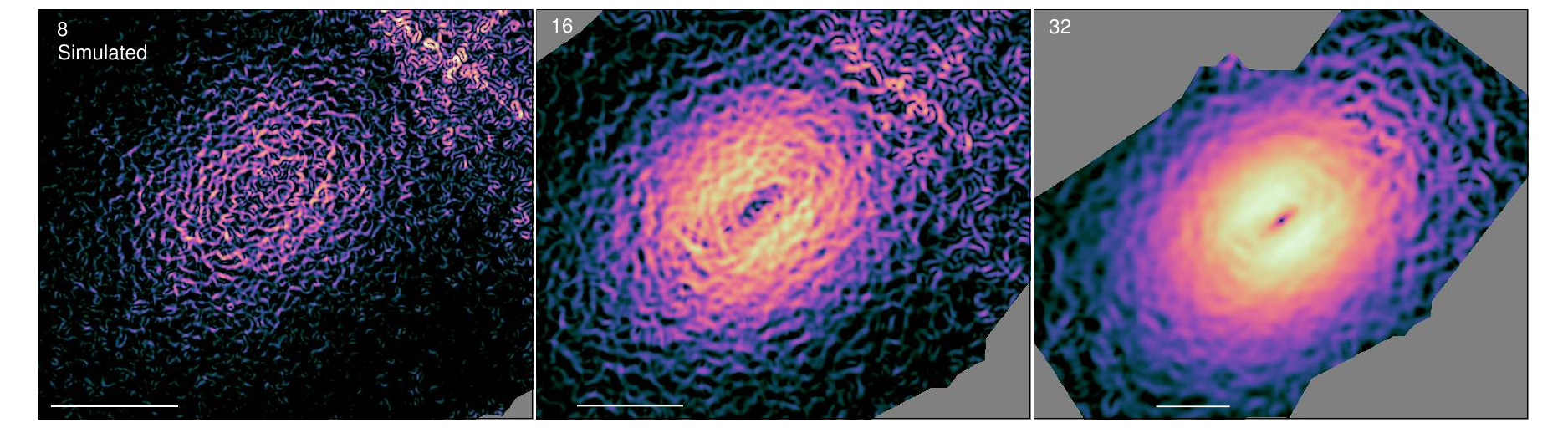}
  \caption{(Top row) GGM-filtered images of A\,3667 with $\sigma=8$,
    16 and 32 pixels (1 pixel is 0.492 arcsec). The bar has a length
    of 4 arcmin (260 kpc).  The smallest and
      largest regions here are shown in Fig.~\ref{fig:expcorr}. The
      white box shows the region examined in the surface brightness
      and gradient profiles. (Bottom row) Filtered images of a
      Poisson realisation of an elliptical $\beta$ model fitted to the
      X-ray data, showing the large-scale gradient and noise.}
  \label{fig:a3667_scales}
\end{figure*}

\subsubsection{M\,87}
The X-ray-emitting jet in M\,87 \citep{Marshall02,WilsonYang02} is
clearly seen at the centre of the smallest scale map
(Fig.~\ref{fig:m87_scales}). Surrounding the jet lies a cocoon of
relativistic plasma which creates cavities in the X-ray emission in
the jet and counter-jet directions \citep{Young02,Forman07}. In the
$\sigma=4$ map are the two well-known arms of soft X-ray emission
\citep{Bohringer95} extending out from the cluster along the arms of
the radio source, likely metal rich material lifted by the radio
bubbles \citep{Young02,Simionescu07,Million10}.  The inner 10 kpc
contains a series of bubbly structures, particularly along the eastern
arm. The arm may be made up of a series of buoyant bubbles. In the
eastern arm the radio plasma and X-rays are cospatial, while along the
southwestern arm they are anticoincident \citep{Forman07} and appear
to spiral around each other \citep{Forman05}. Surrounding the nucleus
with a radius of $\sim 13 \kpc$ is a circular structure
\citep{Young02,Forman05} which is a weak shock \citep{Forman07}. The
$\sigma=8$ pixel map shows that there are a series of edges which have
a similar curvature to the shock, particularly in the region towards
the southwest. There appears a second edge beyond the shock towards
the northeast. We also see linear structures where the arms cross the
shock. On the largest scales are outer cavities in the X-ray emission,
surrounded by an other ring \citep{Forman07}.

\begin{figure}
  \includegraphics[width=\columnwidth]{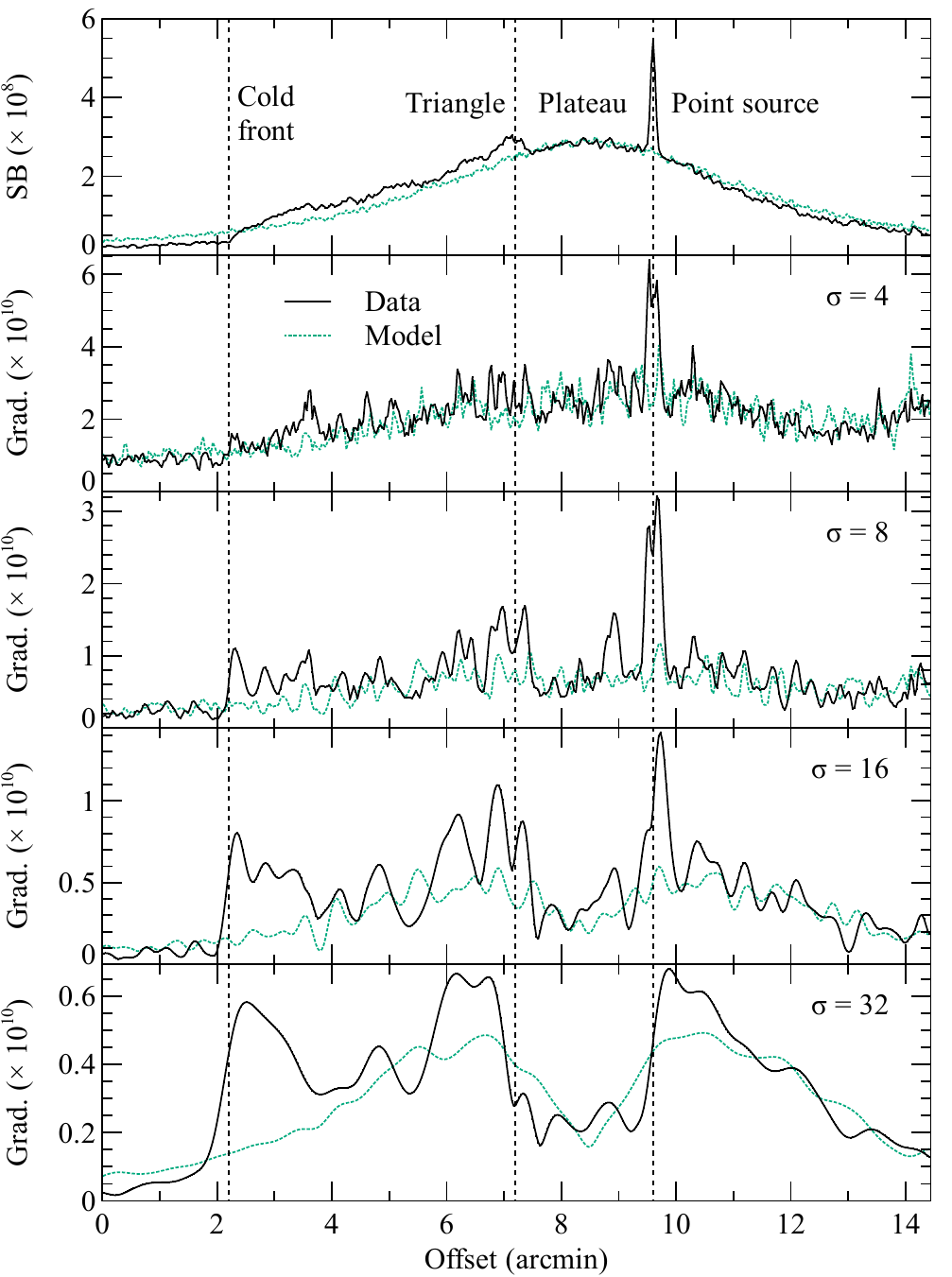}
  \caption{Surface brightness and gradient profiles in a strip across
    A\,3667 from the south-west to the north-east
    (Fig.~\ref{fig:a3667_scales}). The X-ray surface brightness in
    $\photoncmsqps$ is shown in the top panel. The other panels show
    the gradient magnitude for the scales shown in $\photoncmsqps \:
    \mathrm{pixel}^{-1}$, for 0.492 arcsec pixels.
    For comparison are plotted the surface brightness and
    filtered profiles of a Poisson
    realisation of a model (Fig.~\ref{fig:a3667_scales} bottom row).}
  \label{fig:a3667_prof}
\end{figure}

\subsubsection{A\,3667}
There is less dynamic range in the image of A\,3667 and so we only
display the filtered maps for three spatial scales
(Fig.~\ref{fig:a3667_scales} top row). Despite this, the filtering
enhances a number of structures, many of which are not obvious in the
original image. To help assess the significance of
features we show simulated images of the cluster in
Fig.~\ref{fig:a3667_scales} (bottom row). These are based on an
elliptical $\beta$ model fit to the X-ray data, which has no
structure except for a central core. A Poisson realisation of the
fit was filtered in the same way as the original data, showing the
overall gradient and noise.
To examine the significance of the features we calculated
profiles along the main axis of the cluster in surface brightness
and GGM-filtered images on different scales
(Fig.~\ref{fig:a3667_prof}), including the $\sigma=4$ scale. For
comparison we plot the profiles from the simulated model, which
reproduces the overall gradient profiles and level of noise. There
are edges seen in the data on all filtering scales which are not
seen in the filtered model.  There are significant sharp structures
seen in the $\sigma=4$ and 8 profiles which become washed out or
mixed with other structures in the larger scale maps. On larger
scales, the gradient filter reveals longer scale gradients which are
easily missed in the noise in the smaller scale maps.

The most prominent structure is the well-known cold front indicating
that the bright X-ray emitting region is moving through its
surroundings at approximately sonic speeds \citep{Vikhlinin01}. In
these deeper data we see that the edge is not perfectly smooth but
there are features along it, likely Kelvin-Helmholtz
instabilities. There is some faint X-ray emission associated with the
cD galaxy, but a bright point source lies 20 arcsec (22~kpc) to the
south-west. We were not able to fully remove this
point source given its brightness, size and complex surrounding
structure. To the south-east of the galaxy are bright well-defined
blobs, clearly seen in the original X-ray image,
one of which is a triangle-shaped region approximately 60~kpc in
size. These regions are lower in temperature in
their surroundings and have sharp edges. They appear to be in rough
pressure equilibrium with their surroundings. They could be material
stripped during the merger, but it is unclear where the material was
stripped from.

Running between the central galaxy and the triangle is a relatively
featureless region, labelled the inner plateau. Surrounding this
plateau are many linear structures, in particular behind the cold
front. These could be projected Kelvin-Helmholtz
  instabilities (see fig.~8 of \citealt{Roediger13}). We see the
edges of a large scale excess and depression, claimed to be
300-kpc-long Kelvin-Helmholtz instabilities \citep{Mazzotta02}. At the
north-west of our field, where the data quality is poorer, are long
linear edges (labelled Outer Edges). These structures can also be seen
in the deep \emph{XMM} data of \cite{Finoguenov10}.
They are not coincident with the edge of the radio
relic to the north of the cluster, but lie a few hundred kpc inside
them. They could mark the edge of a stripped tail of material, more
easily seen in the \emph{XMM} data.  To the south of the cold front
is another edge, likely the edge detected by \cite{Vikhlinin01}, which
was identified by them as a possible bow shock.

\subsection{Detecting gradients in Poisson noise images}
\label{sect:detgrads}

The ability of the GGM filter to detect a gradient in surface
brightness depends on the magnitude of the jump, the length it occurs over,
the surface brightness and
the value of $\sigma$. As the GGM filter computes a gradient
magnitude, the output must always be zero or positive. Poisson
fluctuations in the input image will therefore produce a noise signal
in the output which is not removed by averaging over area.

\begin{figure}
  \includegraphics[width=\columnwidth]{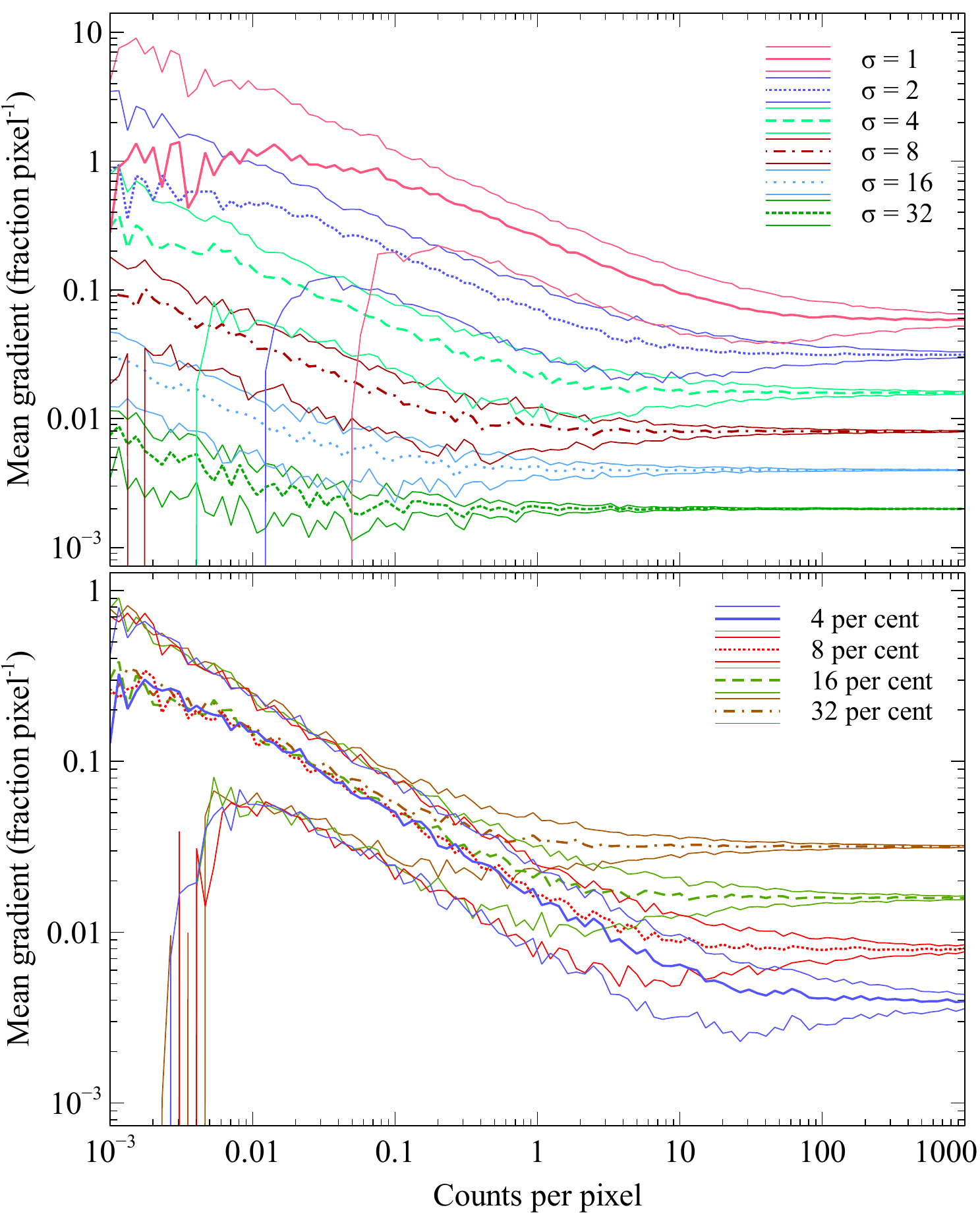}
  \caption{Fractional gradient value detected for a
    1D sharp jump in surface brightness.  The lines show the mean
    values and standard deviations as a function of surface
    brightness. (Top panel) Values using $\sigma=1$, 2, 4, 8, 16 and
    32 pixels for a 16 per cent jump. (Bottom panel) Values for jumps
    of 4, 8, 16 and 32 per cent using $\sigma=4$ pixels.}
  \label{fig:jump_grads}
\end{figure}

To assess this quantitatively, we repeatedly filtered Poisson realisations of a
simple model images with a one-dimensional jump in surface brightness
with zero width around a certain mean surface
brightness. Fig.~\ref{fig:jump_grads} (top panel) shows how the
resulting gradient value at the jump pixel
varies as a function of surface brightness
and $\sigma$ for a fixed fractional surface brightness jump.
At low count rates the gradient signal becomes
increasingly dominated by noise. At higher count rates the mean gradient
tends towards a constant value and the standard deviation decreases.
The noise signal at low count rates has the same Poisson error origin
as the noise in the real signal at high count rates, scaling as the
surface brightness to the power $-1/2$.  This noise component also
scales as $\sigma^{-1/2}$, as expected if sensitive to
the number of counts within the filter.
As the filter has a finite width and the jump
is narrow, the determined gradient varies as $1/\sigma$.
If a model with continuous gradient over a few $\sigma$ is examined,
its value is recovered for all values of $\sigma$.

The ability to detect a jump depends on the magnitude of the jump, shown
using $\sigma=4$ in Fig.~\ref{fig:jump_grads} (lower panel). Jumps become
more visible with increasing jump size and count rate.
In this one-dimensional case the count rate at which the gradient diverges
from the noise profile decreases
with the inverse of the fractional magnitude of the jump.

\subsection{Assessing the significance of structures}

\begin{figure}
  \centering
  \includegraphics[width=0.9\columnwidth]{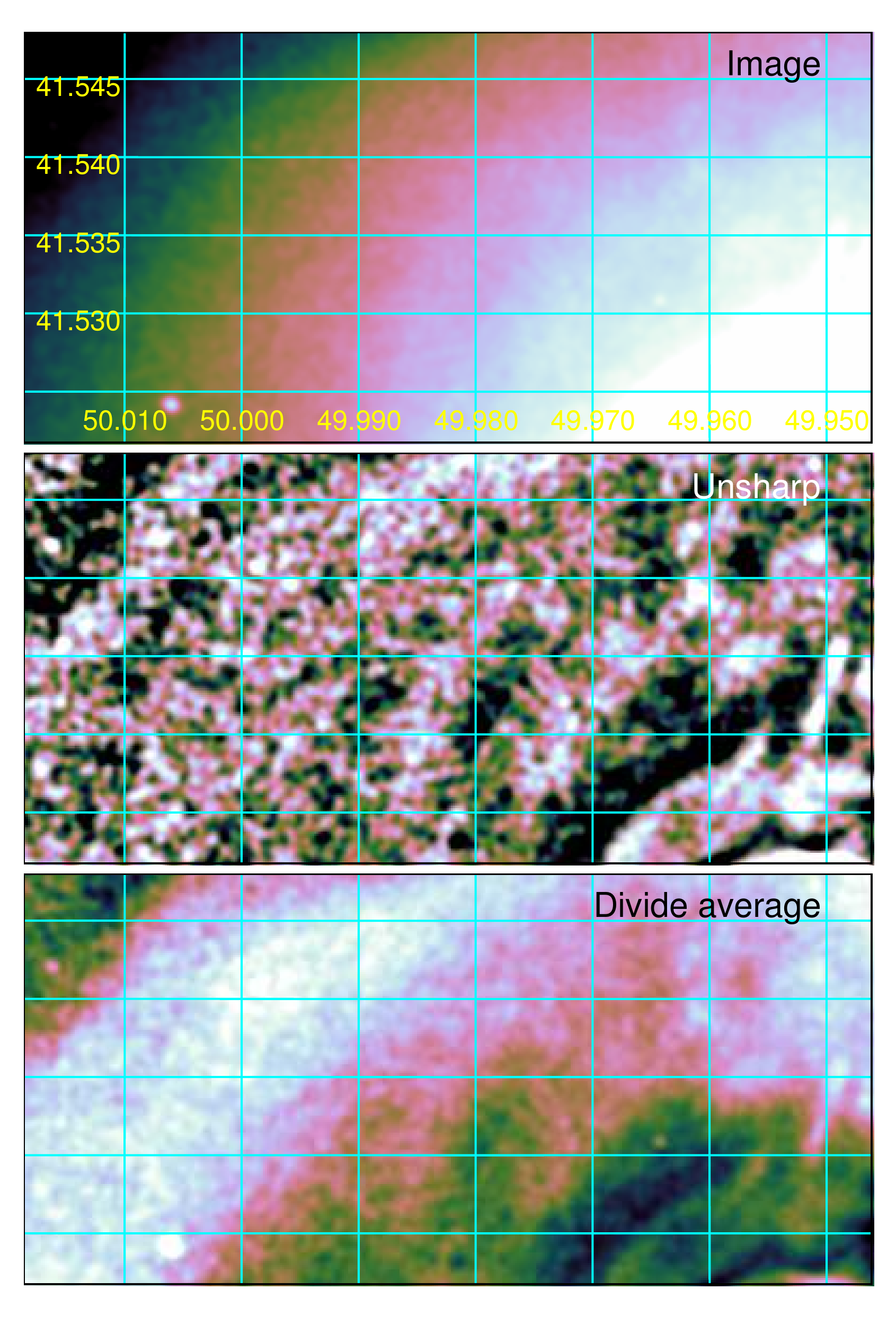}
  \includegraphics[width=0.9\columnwidth]{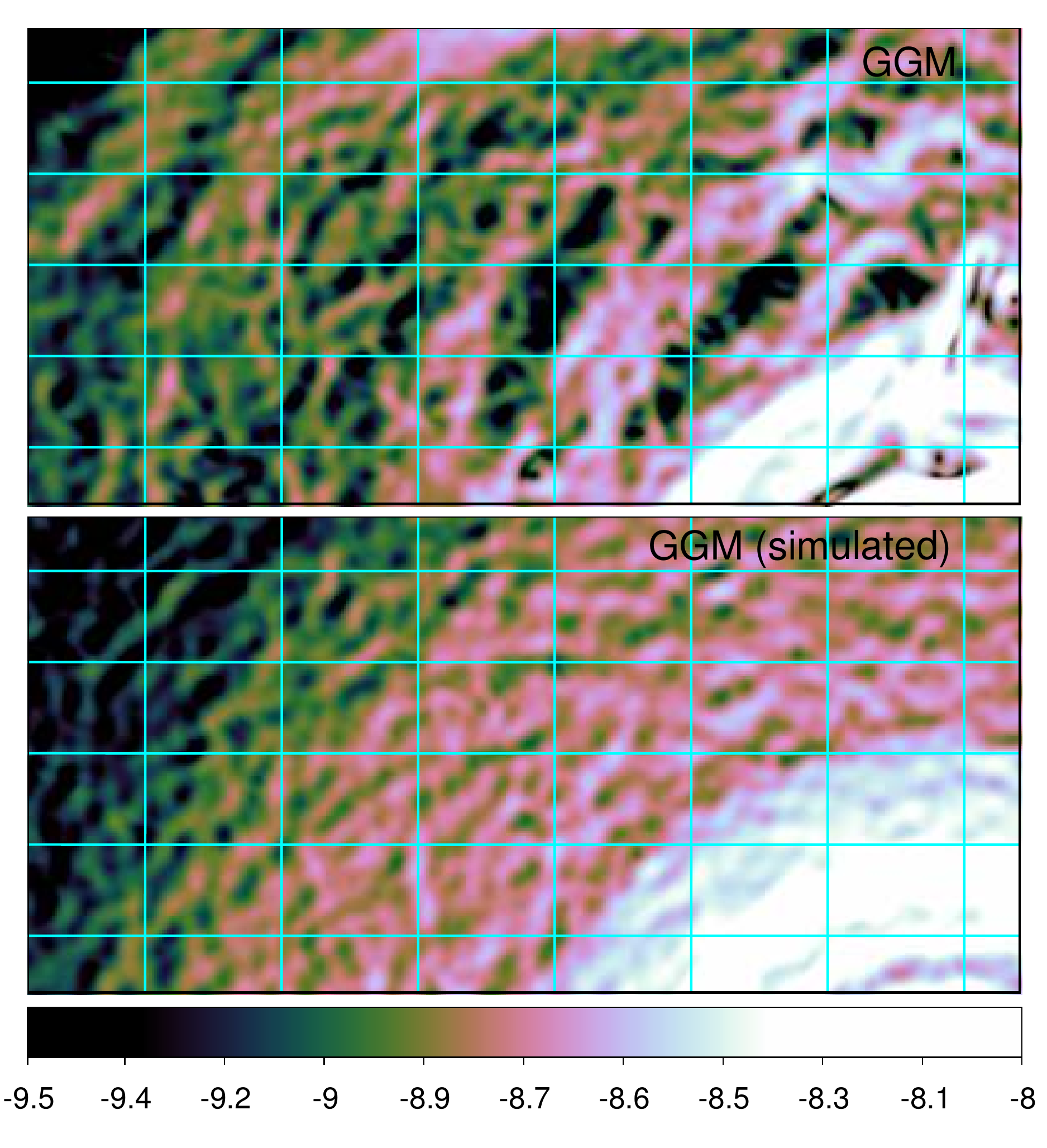}
  \caption{Comparison between different methods applied to the same
    region of Perseus. Top panel: X-ray image, smoothed by a
    $\sigma=2$ Gaussian.  2nd panel: unsharp-masked image, dividing
    Gaussian-smoothed images with $\sigma=2$ and $16$ pixels. 3rd
    panel: ratio between data and average at each radius, smoothed
    using $\sigma=2$. 4th panel: GGM-filtered image ($\sigma=4$,
    hiding point sources), with scale in $\log_{10} \photoncmsqps
    \pixel^{-1}$, 5th panel: GGM-filtered simulated data with the
    average radial surface brightness as the real data and same colour
    scale as 4th panel.}
  \label{fig:comparereg}
\end{figure}

Although GGM filtering is a simple image analysis technique, it is
important to assess the significance of features in resulting
maps. Our ability to detect gradients depends on the
  count rate, size of gradient (in both magnitude and length) and
  $\sigma$ (Section \ref{sect:detgrads}). The simplest method to
assess significance is to compare the filtered image with raw
data. Often structures can be directly observed in the raw data by
using a colour scale well matched to the region in question, or by
blinking between the filtered and original images.
The significance of features can also be assessed by comparing them
with the strength of noise at the same radius, where the count rate
is likely to be similar.  Filtered images can also be compared to
unsharp-masked images or images with the radial average removed
to ensure that features are robust. A further
technique is to make a simulated cluster image, based on a smooth
surface brightness profile or other model, and to filter this in the
same way as the data. The real and simulated filtered images can then
be compared to assess the significance of structures
(see e.g. Figures \ref{fig:a3667_scales} and
\ref{fig:a3667_prof}).  Comparison of images with those at other
wavelengths can also confirm the existence of structures in filtered
images (see Sections \ref{sect:scales} and \ref{sect:shocks}).

We examine a small region to the north-east of the Perseus cluster
core (Fig.~\ref{fig:comparereg}), showing a smoothed X-ray image, a
GGM-filtered image, a filtered image of a simulation of the cluster
with the same radial profile, an unsharp masked image and an image
showing the fractional residuals to the average at
each radius. The magnitude of the structures in the real data is
much larger than those in the simulated filtered map, although we see
that the real data has noise in it at a similar level to the filtered
map. The noise patterns tend to be small linear structures which lie
perpendicular to the surface brightness gradient. The main gradient
component that the filter is measuring is radial and so fluctuations
due to noise are most strongly seen in the radial direction after
filtering, leading to these characteristic noise patterns. A key
method for assessing the level of noise is to look at the amount of
noise in other directions where the gradient and data quality are
similar.

The A\,3667 data (Fig.~\ref{fig:a3667_scales}) highlight that the
appearance of noise after the filtering process depends on the data
quality. In the north west parts of the 8 and 16 scale images there
are regions filled with fluctuating dark and bright structures. This
is a region in the cluster which is only covered using relatively
short exposures. Therefore one must be careful when assessing features
in filtered data where the data quality varies strongly over the
image.

\subsection{Combining images of different scales}
\label{sect:scales}

\begin{figure}
  \includegraphics[width=\columnwidth]{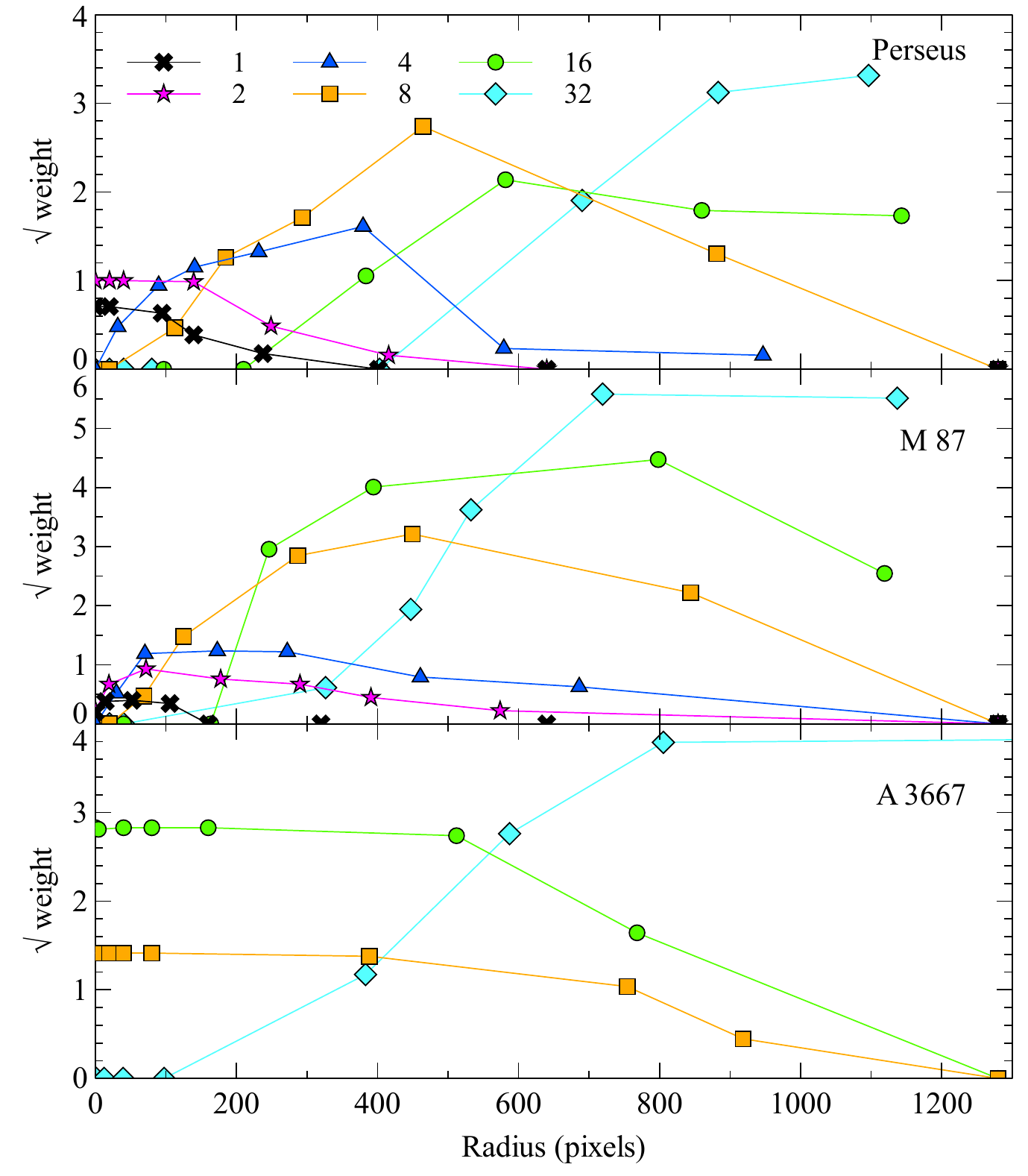}
  \caption{Radial profiles of the radial weights used when making the
    combined filtered images of the Perseus cluster, M\,87 and
    A\,3667.}
  \label{fig:scalefactors}
\end{figure}

Filtering on a certain length scale is applicable to a particular
region in the cluster, where the count rate is large enough to allow
the gradient to be measured. We have therefore implemented a scheme
where we add the images with different scales together, weighting each
of the images using a radial weighting scheme. We adjust a number of
control points (radius and weighting factor) and use linear
interpolation to calculate the values for intermediate radii. The
motivation for the radial weighting is that the number of counts, the
main quantity which determines our ability to measure the gradient,
mainly varies radially in clusters. The weighting procedure also
allows the magnitude of the features in the centre to be suppressed
relative to the outskirts, in order to plot them on the same image. We
constructed a graphical user interface in order to adjust the
different radial scaling factors. Fig.~\ref{fig:scalefactors} shows
the relative radial weights of the different filtered images for
the three clusters.

\begin{figure*}
  \includegraphics[width=\textwidth]{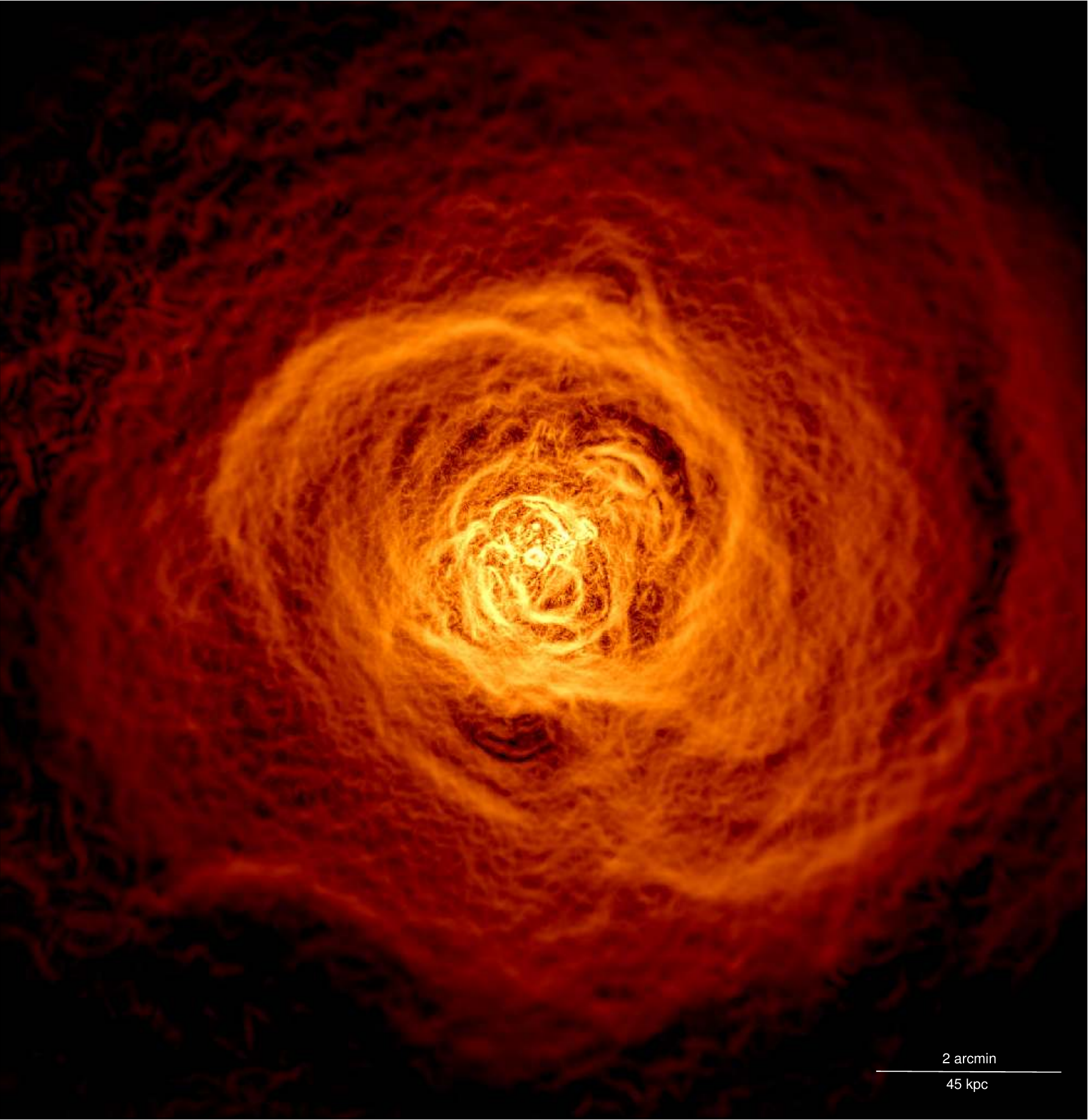}
  \caption{Combined GGM-filtered of the Perseus cluster, adding maps
    with $\sigma=1$ to 32 pixels with radial weighting.}
  \label{fig:per_combine}
\end{figure*}

\begin{figure*}
  \includegraphics[width=\textwidth]{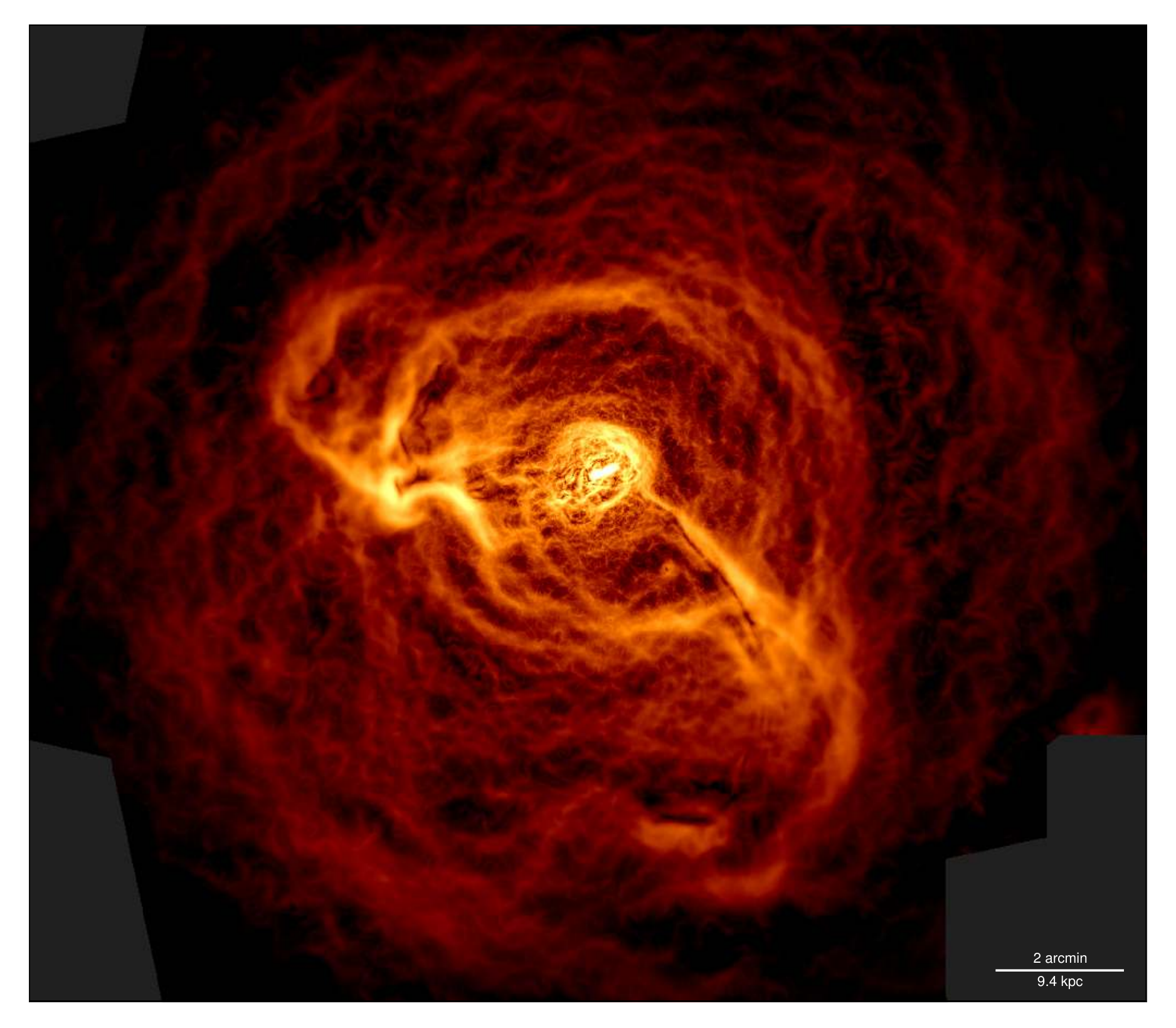}
  \caption{Combined GGM-filtered of M\,87, adding maps with $\sigma=1$
    to 32 pixels with radial weighting.}
  \label{fig:m87_combine}
\end{figure*}

Fig.~\ref{fig:per_combine} shows the combined image for the Perseus
cluster. In the inner part of the cluster, most of the signal comes
from the $\sigma=1$ and $2$ maps, while in the outskirts the
$\sigma=8$, $16$ and $32$ maps are
combined. Fig.~\ref{fig:m87_combine} shows the combined results for
M\,87.

\begin{figure}
  \includegraphics[width=\columnwidth]{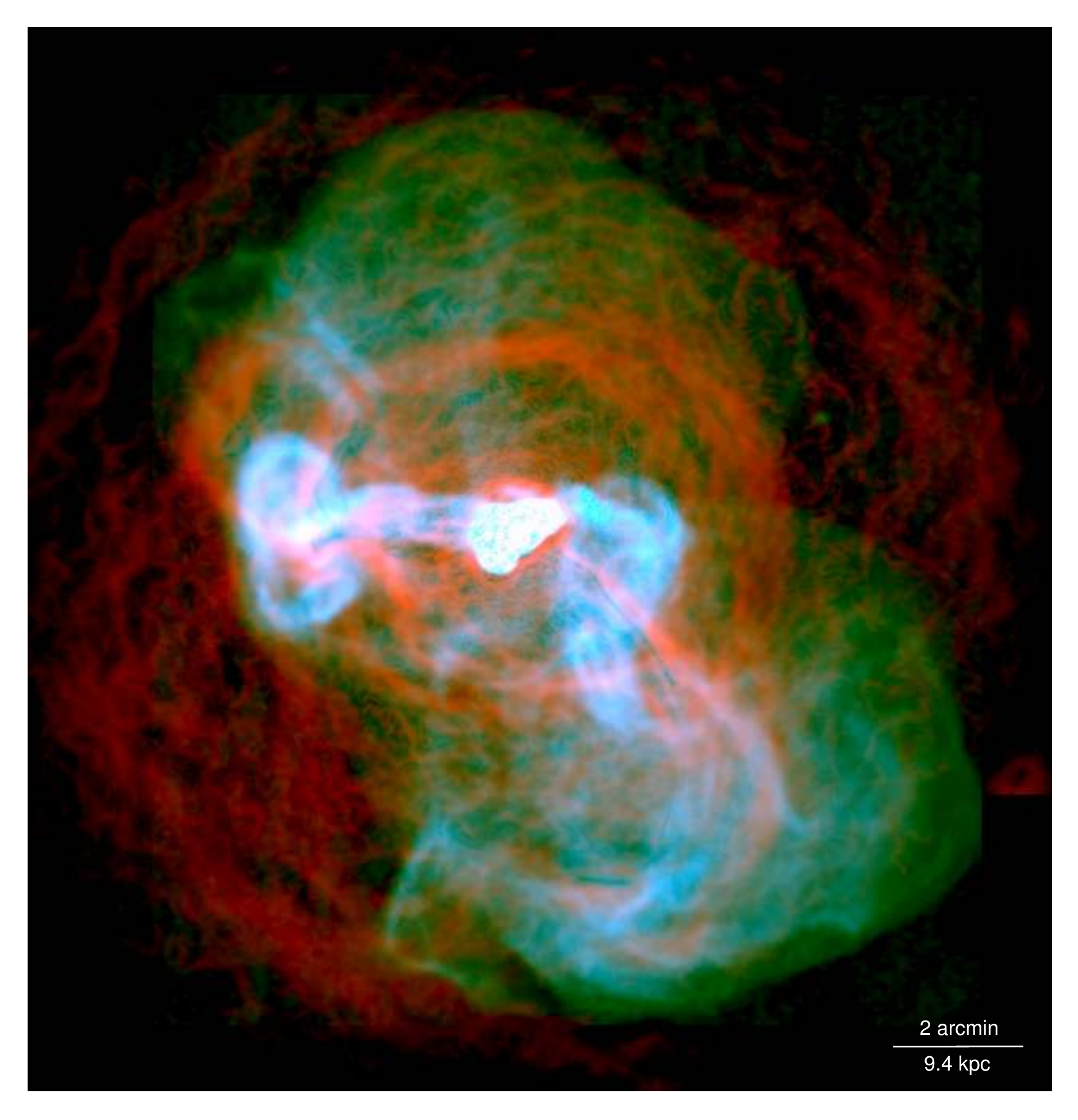}
  \caption{GGM-filtered X-ray image of M87 from
    Fig.\ref{fig:m87_combine} (red), overlayed with 90~cm radio
    emission from \protect\cite{Owen00} (blue/green).}
  \label{fig:m87_radio}
\end{figure}

Fig.~\ref{fig:m87_radio} compares the filtered X-rays with
the 90~cm radio emission \citep{Owen00}, highlighting the close
connection between the X-ray and radio-emitting plasmas.
The radio source has
two arms, one to the east with a `mushroom' appearance and the other
to the south west with a filamentary structure. The southwestern arm
has a twisted appearance, where the X-rays and radio appear to be
anticoincident, likely dominated by magnetic structures \citep{Forman07}.
The eastern arm, in contrast, has coincident X-ray and radio structures.
This arm may be made up of a series of small radio bubbles plus the
large radio torus which makes the cap of the mushroom \citep{Forman07}.
Much of the edge of large scale radio structure is coincident with edges
in the X-ray surface brightness.

\begin{figure}
  \includegraphics[width=\columnwidth]{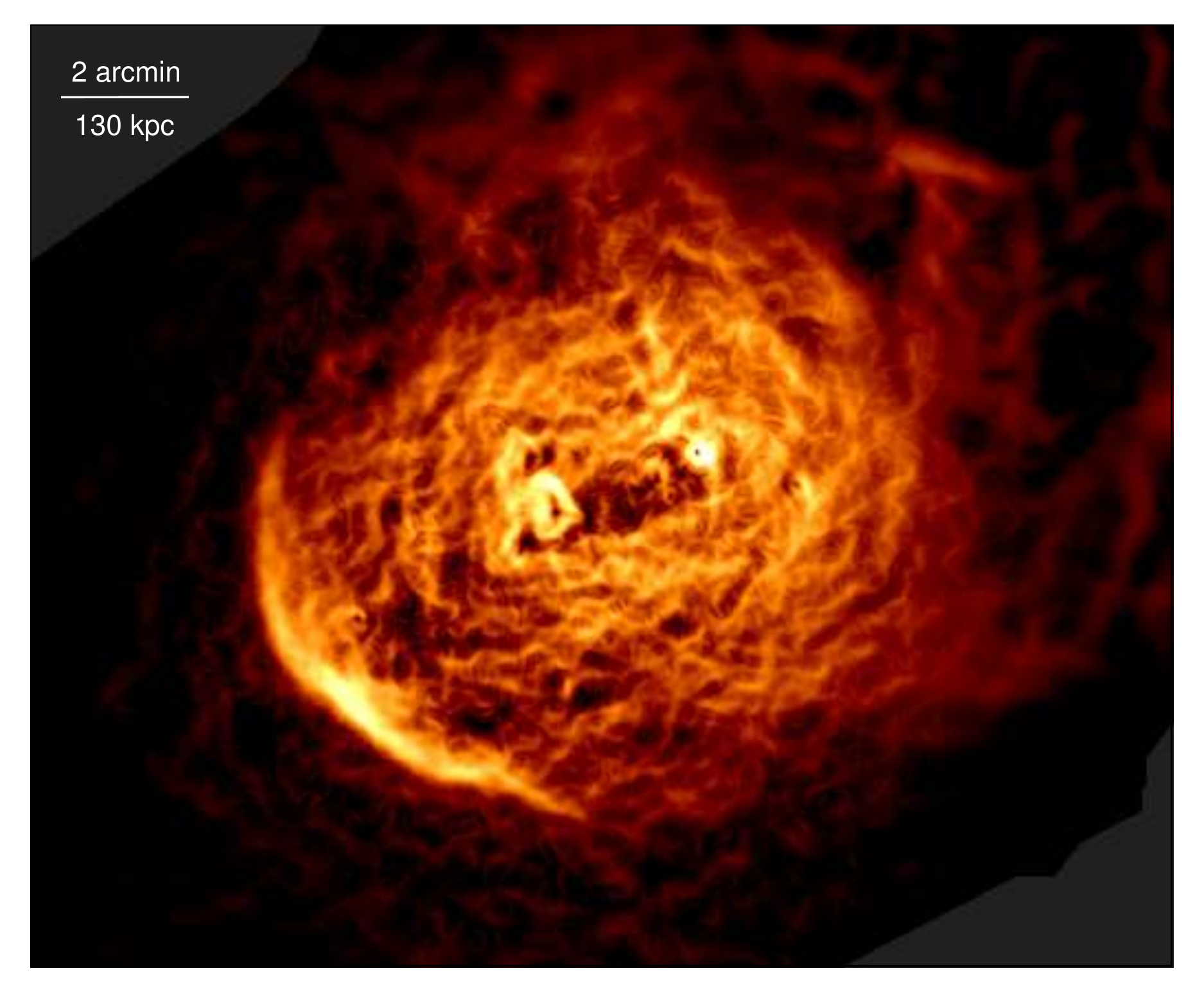}
  \caption{Combined GGM-filtered image of A\,3667, adding maps with
    $\sigma=8$, 16 and 32 with radial weighting.}
  \label{fig:a3667_combine}
\end{figure}

Fig.~\ref{fig:a3667_combine} shows a map combining filtered images
with scales of $\sigma=8$, 16 and 32 pixels for A\,3667. It highlights
the sharp cold front edge and surface brightness plateau between the
triangle structure and the central galaxy. The combined multi-scale
images demonstrate that it is possible to create useful qualitative
maps which show the majority of the surface brightness edges in a
galaxy cluster image.

\subsection{Comparison with other methods}
The GGM-filtered image of the example region
(Fig.~\ref{fig:comparereg}) clearly reveals the surface brightness
edges in the data, particularly those associated with the ripple-like
structures.

These ripples are not so easily seen in the image
showing the fractional deviations from the radial average, likely
due to them being mostly azimuthal in morphology. When subtracting
models as here, the choice of a model can greatly influence the
produced residual image \citep{SandersAWM712}, so care must be taken
to not introduce or remove important structures.  It is difficult to
construct models which replicate the complex structure seen, for
example, as the spiral in the core of the Perseus cluster.

Unsharp masking does show the ripples here, but they have a relatively
unclear and patchy appearance. This is similar to the results of an
analysis of images of the Centaurus cluster, where the linear
structures present are revealed using GGM filtering, but not using
unsharp masking \citep{SandersCent16}. A disadvantage of gradient
filtering over unsharp masking, however, is that intrinsically narrow
filamentary structures are broadened, such as the filament to the
north (on the western side of the image). A filament is converted to a
double structure due to the sharp gradient on either side and the flat
gradient along its ridge.

\begin{figure}
  \includegraphics[width=\columnwidth]{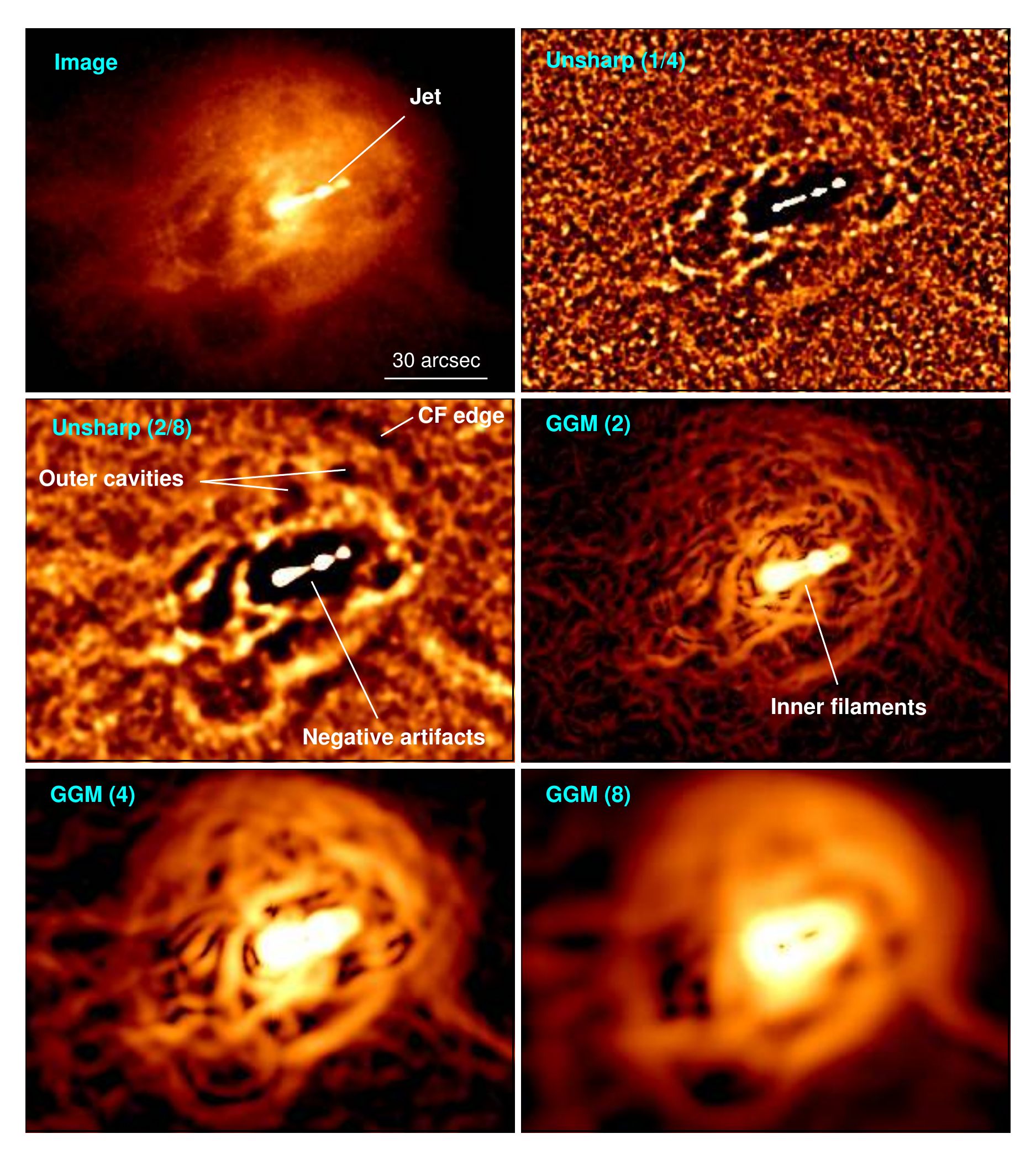}
  \caption{Comparison of unsharp masking and gradient filtering in the
    central region of M\,87. Top left panel: X-ray image with point
    sources removed and smoothed by a Gaussian of $\sigma=1$
    pixel. Top right panel: unsharp masking, showing fractional
    difference between images smoothed by 1 and 4 pixels. Centre left
    panel: unsharp masking using 2 and 8 pixels. Other panels:
    gradient filtered image with $\sigma$ as value given in pixels.}
  \label{fig:m87_compare}
\end{figure}

To compare unsharp masking and gradient filtering in more detail, we
show in Fig.~\ref{fig:m87_compare} the highly-structured central
region of M\,87. Shown are the smoothed data, two unsharp-masked
images (using different large and small smoothing scales) and three
GGM-filtered images (using scales of 2, 4 and 8 pixels). Unsharp
masking commonly produces negative artifacts surrounding bright
sources (seen near the jet here) which are difficult to distinguish
from, or mask, AGN-generated cavities in the ICM
(labelled negative artifacts).
Immediately surrounding the jet are filamentary structures
(labelled inner filaments) visible in the GGM-filtered images which
are lost in the unsharp-masked data, due to the large negative
residual artifacts there.

Unsharp masking also produces similar negative artifacts at sharp
surface brightness edges such as shocks or cold fronts, seen here to
the north-west (labelled CF edge).
GGM-filtering does not produce similar signals which
could otherwise be misinterpreted as cavities. The negative residuals
around cold fronts in unsharp-masked images are an issue when trying
to detect cavities in more distant objects, where there can be large
changes in surface brightness. Unsharp masking relies on a smoothed
image being a good approximation for the underlying cluster emission,
which is likely not to be the case in the peaked central regions.
Some features are however clearer in the unsharp-masked images,
particularly those which are smaller than the smoothing scale on a
relatively flat background, for example the small depressions to the
north of the image (labelled outer cavities).
Nevertheless, the GGM-filtered $\sigma=2$ and $4$
images reveal a great deal of structure which is not obviously present
in the unsharp-masked data. It is clear that continuous structures are
better connected on this complex underlying cluster emission in the
GGM-filtered images than when using unsharp masking.

Despite the power of the gradient filtering method, there are relative
advantages and disadvantages of GGM filtering, unsharp masking and
model subtraction. Therefore, a combination of the various techniques is
likely to help reliably identify structures, particular at higher
redshifts.

\subsection{Logarithmic gradients}
The average surface brightness profile for a galaxy cluster can be
approximated by a powerlaw in radius, at least over large radial
regions. Ideally to identify where there are deviations from a smooth
profile, it would be better to examine the gradient of the logarithm
of the surface brightness in logarithmic radius, rather than in linear
coordinates. We can partially achieve this by computing the logarithm
of the surface brightness and applying a gradient filter to the
resulting image. As X-ray images are noisy and there are often pixels
with zero values, a better way of doing this is to first smooth the
X-ray image with the Gaussian of the scale required, take the
logarithm of the smoothed image, and then compute the pixel-by-pixel
gradient magnitude.

\begin{figure}
  \includegraphics[width=\columnwidth]{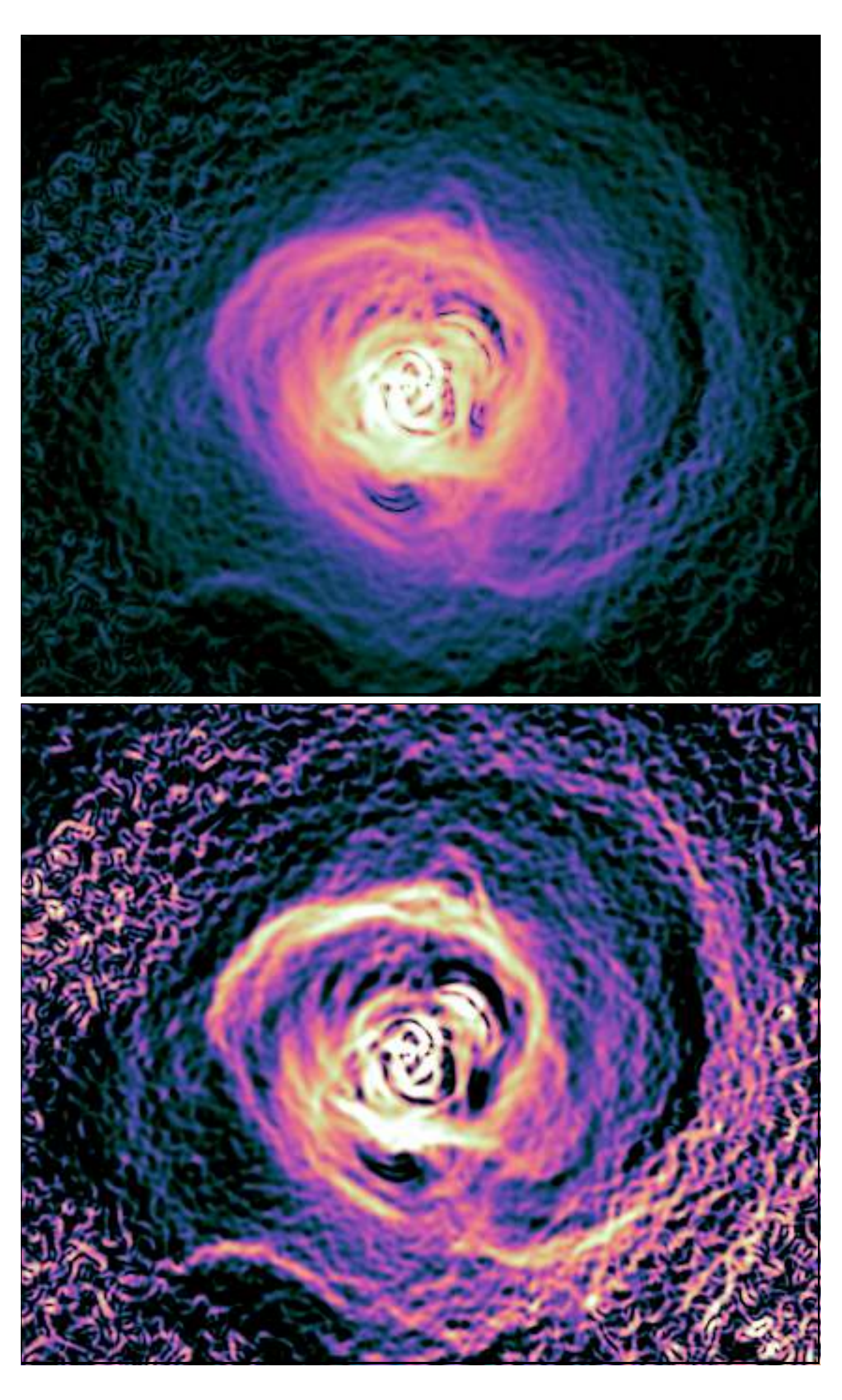}
  \caption{Gradient image of Perseus ($\sigma=8$ pixels) (top panel)
    compared to a gradient of the log value of an image smoothed using
    a Gaussian of $\sigma=8$ pixels (bottom panel).}
  \label{fig:perlog}
\end{figure}

This method can produce good results, as shown by the filtered image
of Perseus in Fig.~\ref{fig:perlog}. The component of the gradient
which comes from the cluster profile is significantly reduced. The
image shows the structure from the inner shock out to the large scale
spiral using a single filtering scale.

A disadvantage of the technique is that the value becomes noisy where
the count rate is low in the outskirts. When the gradient of a
non-logarithmic X-ray image is computed, the pixels with low number of
counts typically are in regions with low absolute gradients and so the
noise on the gradient is low compared to the gradient value in the
centre where the counts rate are high. However, using a logarithmic
image the gradient in the outskirts is similar to the value in the
centre. As the count rate in the outskirts is lower, the scatter in
the value is higher. This noise can be seen in the north-east and
south-west parts of Fig.~\ref{fig:perlog}, where the observation is
shallower and the cluster fainter.

Logarithmic gradient images are therefore likely preferred if there
is a sufficiently high count rate across the region of interest
so that the gradient can be measured to a high fractional accuracy.
However, this criterion is unlikely to be met using typical
photon-starved observations with \emph{Chandra} except in the core
region or using large spatial scales.

\subsection{Finding shocks}
\label{sect:shocks}
As pointed out by \cite{Forman07}, between temperatures of around 1
and 3 keV the \emph{Chandra} 3.5 to 7.5 keV band count rate is
approximately proportional to the pressure-squared integrated along
the line of sight. Therefore, by gradient filtering such images, we
are able to detect pressure discontinuities and shocks in clusters.

\begin{figure*}
  \includegraphics[width=\textwidth]{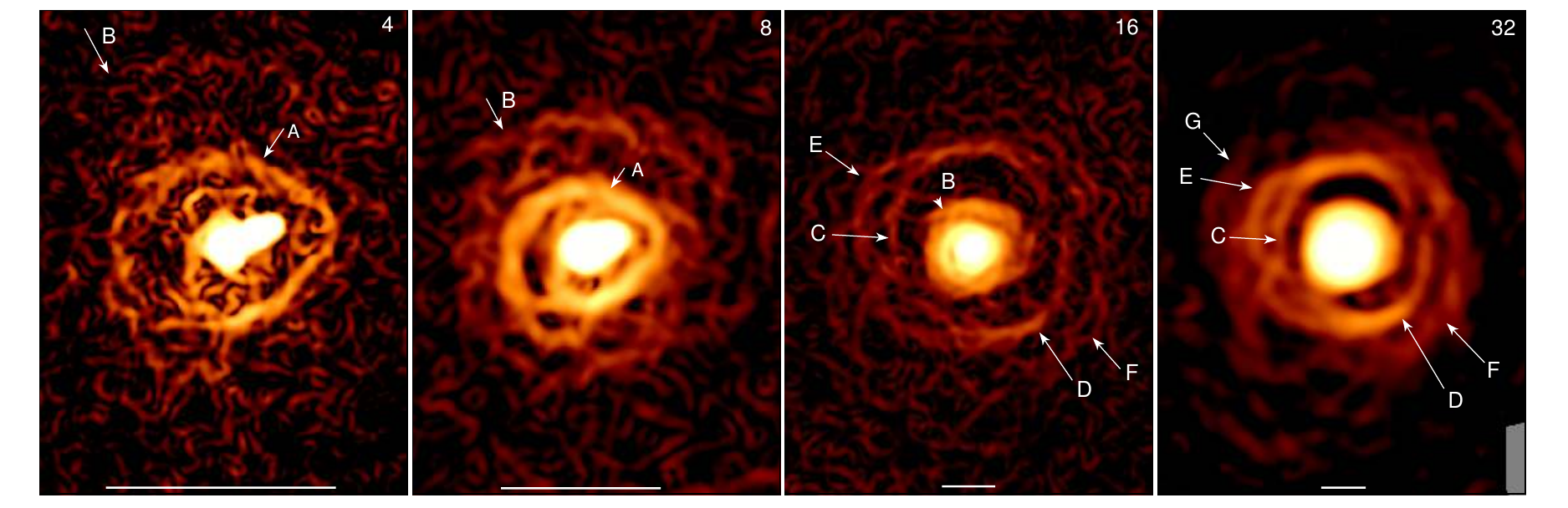}
  \caption{Pressure-jump-sensitive GGM-filtered hard-band 3.5 to 7.5
    keV images of M\,87.  The images use $\sigma=4$ to 32 pixels,
    showing a set of pressure discontinuities marked by arrows.  The
    bar has a length of 1.5 arcmin (7.1 kpc).}
  \label{fig:m87_hard}
\end{figure*}

Fig.~\ref{fig:m87_hard} shows filtered images from scales of 4 to 32
pixels of M\,87 in this hard X-ray band. In the centre is an
egg-shaped region previously identified by \cite{Young02}, marked by A
in the 4 and 8 scale maps. This is likely a high pressure region
created by the current AGN outburst. Surrounding this feature is a
second edge in pressure (labelled B), seen clearly
in the 8 map and at lower significance in the 4 map. At a radius of
13~kpc is the clearest shock (labelled C and D), believed to be driven
by an earlier AGN episode approximately 14~Myr ago \citep{Forman07}.
Spectral fitting shows it to have a Mach number of
1.25 \citep{Million10}.

The most interesting aspect of this image, in agreement with the full
band image and filtered image, is that the 13~kpc shock is not a
complete circle, but breaks up into multiple edges (C, D and E), with
a further edge at lower surface brightness levels
(F). Edges F and G are stronger than the noise
at the same radius and can be seen in the unfiltered image. F lies
at the edge of the south-west radio plume. The splitting up of the C,
D, and E structure may be due to varying temperature structure along
the line of sight, affecting the sound speed. Alternatively there
could have been multiple outbursts.

\begin{figure}
  \includegraphics[width=\columnwidth]{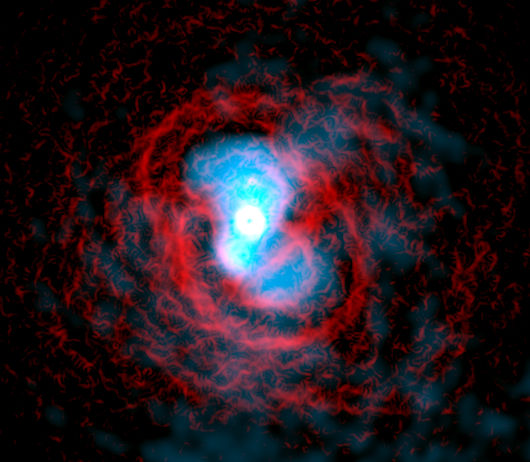}
  \caption{Filtered 3.5 to 7.5 keV X-ray image of Perseus (red) with
    330~MHz radio emission (blue; \citealt{FabianPer02}). The X-ray
    image is a linear combination of the $\sigma=2$, 4 and 8 pixels
    maps. The image measures 4.3 arcmin (96 kpc) across.}
  \label{fig:per_hard}
\end{figure}

In the Perseus cluster the intracluster medium is too hot for these
hard-band images to be solely sensitive to pressure
variations. However, in the central region around the inner cavities
the 2.7 to 4 keV temperatures are close to the preferred
range. Fig.~\ref{fig:per_hard} shows a filtered hard X-ray image of
Perseus (combining three different scales), overlaying the radio
emission. The image highlights the jumps in pressure at the edge of
the shocks surrounding the inner cavities. The image can be
interpreted as two shocks, one surrounding each bubble (as seen in a
spectral fitting pressure map; \citealt{FabianPer06}). The southern
rim of the northern shock appears to pass through the southern cavity
and the rim of the northern rim of the southern shock appears to pass
through the northern cavity. The straight feature across the southern
cavity appears to be the edge of the shock and is not related directly
to the radio source. To the north-west the outer edge of the shock is
much less clear, where the radio plasma extends from the northern
inner cavity to the outer north-western ghost cavity.

\section{Conclusions}
We examine X-ray images of the Perseus cluster, M\,87 and A\,3667 with
the Gaussian gradient magnitude filter to detect edges. We show that
the filter is able to detect a host of structures within these
clusters. The method is often more sensitive to features than
unsharp-masking or subtracting radial cluster models. It also does not
introduce negative residual artifacts commonly seen in unsharp-masked
images.  By the use of a radial weighting scheme we can produce a
multi-scale image which demonstrates that a wealth of physical
processes are occurring in these clusters. Using pressure-sensitive
hard-energy-band images it is possible to use the method to detect
shocks in clusters.

\section*{Software repository}
The code described and used in this paper can be found at
\url{https://github.com/jeremysanders/ggm}.

\section*{Acknowledgements}
ACF, HRR and SAW acknowledge support from the ERC Advanced Grant
FEEDBACK. The scientific results reported in this article are based on
data obtained from the Chandra Data Archive.

\bibliographystyle{mnras}
\bibliography{refs}

\end{document}